\documentclass[a4paper,floatfix,notitlepage,rmp]{revtex4-1}
\usepackage[british]{babel}
\usepackage[latin1]{inputenc}
\usepackage{amsmath,amssymb,mathtools}
\usepackage{graphicx}
\usepackage[sort&compress]{natbib}

\begin{document}

\title[Ellison and Thorpe scales in the ocean]{
  Comparison of Ellison and Thorpe scales from Eulerian ocean temperature observations.
}

\author{Andrea Cimatoribus}
\email{Andrea.Cimatoribus@nioz.nl}
\affiliation{Royal Netherlands Institute for Sea Research, Texel, the Netherlands}
\author{Hans van Haren}
\affiliation{Royal Netherlands Institute for Sea Research, Texel, the Netherlands}
\author{Louis Gostiaux}
\affiliation{Laboratoire de M\'ecanique des Fluides et d'Acoustique, CNRS/Universit\'e de Lyon, Lyon, France}

\begin{abstract}
Ocean turbulence dissipation rate is estimated either by means of microstructure shear measurements, or by adiabatically reordering vertical profiles of density.
The latter technique leads to the estimate of the Thorpe scale, which in turn can be used to obtain average turbulence dissipation rate by comparing the Thorpe scale to the Ozmidov scale.
In both cases, the turbulence dissipation rate can be estimated using single vertical profiles from shipborne instrumentation.
We present here an alternative method to estimate the length scale of overturns by using the Ellison length scale.
The Ellison scale is estimated from temperature variance just beyond the internal wave band, measured by moored instruments.
We apply the method to high resolution temperature data from two moorings deployed at different locations around the Josephine seamount (North Eastern Atlantic Ocean), in a region of bottom-intensified turbulence.
The variance of the temperature time series just above the internal wave frequency band is well correlated with the Thorpe scale.
The method is based on the time--frequency decomposition of variance called ``maximum overlap discrete wavelet transform''.
The results show that the Ellison length scale can be a viable alternative to the Thorpe scale for indirectly estimating turbulence dissipation rate from moored instruments in the ocean if time resolution is sufficiently high.
We suggest that fine structure contaminated temperature measurements can provide reliable information on turbulence intensity.
\end{abstract}

\maketitle 

\section{Introduction}
\label{sec:intro}

Turbulence dissipation rate is a key quantity in a turbulent flow.
It is defined as $\epsilon=\frac{1}{2} \, \nu \left(\frac{\partial u_{i}}{\partial x_j}+\frac{\partial u_{j}}{\partial x_i}\right)^2$, where $\nu$ is the kinematic viscosity, $u_i$ is the component of the velocity fluctuation along the $x_i$ direction and summation over repeated indices is assumed \citep[see, e.g.,][]{frisch_turbulence_1996}.
The interest in the value of the turbulence dissipation rate is linked to two fundamental open questions on the world ocean circulation.
On one hand, its distribution is essential for better understanding how the kinetic energy budget of the world ocean is closed \citep{ferrari_ocean_2009}.
On the other hand, $\epsilon$ is often used for computing vertical diffusivity, a key quantity controlling processes ranging from nutrient upwelling to the general circulation of the ocean.
This second aspect is further complicated by the lack of agreement on a value, or on a model, for mixing efficiency \citep{ivey_density_2008}.
For this reason we will not consider vertical diffusivity in this work.

Direct estimation of $\epsilon$ in the ocean interior requires the measurement of velocity shear at the millimeter scale where most of the dissipation actually takes place \citep[see for instance the shipborne microstructure measurements of][]{oakey_determination_1982,gregg_diapycnal_1987,polzin_spatial_1997}.
Resolving the microstructure shear in the deep ocean is not routine due to various reasons ranging from the cost of deep-ocean shear profilers to the challenge represented by the interpretation of the raw data due to, e.g., vibrations of the instrument \citep[for a review on the instrumentation, see][]{lueck_oceanic_2002}.

Measurements of shipborne CTD (conductivity, temperature, depth), or temperature and depth alone, are more routinely performed.
Consequently, an indirect approach is often followed enabling the estimation of the average turbulence dissipation rate from the characteristic size of the overturns in a stratified water column \citep{thorpe_turbulence_1977,dillon_vertical_1982,itsweire_measurements_1984,gregg_diapycnal_1987,gargett_ocean_1989}.
In particular, $\epsilon$ scales with the size of the largest overturns unaffected by stratification for a given turbulence level and stratification, the Ozmidov scale $L_O=(\epsilon/N^3)^\frac{1}{2}$, with N the buoyancy frequency.
The Ozmidov scale is not measured directly, but can be linked to an objective measure of the overturn's size in the water column.
A common choice is the Thorpe scale ($L_T$), defined as the root mean square (RMS) of the displacement needed to adiabatically reorder a (CTD) profile containing density inversions due to overturns \cite{thorpe_turbulence_1977}.

\citet{dillon_vertical_1982} estimated the ratio $L_O/L_T$ to be approximately $0.8$.
This linear relationship has been confirmed by several authors like \citet{crawford_comparison_1986,itsweire_turbulence_1993, peters_detail_1995, ferron_mixing_1998} and \citet{stansfield_probability_2001}, in both observations and laboratory experiments, even if the large uncertainty affecting this quantity must be kept in mind (observational estimates of the ratio have errors larger than 0.3).
\citet{yagi_modified_2013}, while reporting a value of the $L_O/L_T$ ratio within the error bars of previous estimates, suggest that the ratio may depend on the measurement location, and should thus be estimated locally.
Other works show that the ratio is not a constant with time \citep[e.g.][]{smyth_length_2000,smyth_efficiency_2001,mater_relevance_2013}.
However, these latter studies consider the time evolution of the $L_O/L_T$ ratio during a single mixing event, while time and spatial averages {have to be} taken in the observations \citep[for a discussion of the essential role of averaging for finding an approximately linear relation between $L_O$ and $L_T$ see][]{peters_detail_1995}.
Keeping in mind these uncertainties connected with the use of Thorpe scales, in this work we will follow \citet{dillon_vertical_1982}, estimating turbulence dissipation rate by the formula:
\begin{equation}
  \epsilon_{L_T} = 0.64\, L_T^2 N^3,
  \label{eq:eps_LT}
\end{equation}
where the subscript of $\epsilon$ is a reminder of the method used for the estimation.
{The buoyancy frequency $N$ is computed as the column average from the reordered density profile.}
If only temperature is measured, as often done in practice {and here too}, a key requirement for applying (\ref{eq:eps_LT}) is that temperature is a good proxy of density.
{In practice, this means that the temperature--density relationship should be well approximated by a single functional form.
A linear temperature--density relationship is a further simplification, in which case density variations can be obtained by using thermal compressibility alone.}

Both estimation methods of the turbulence dissipation rate, direct and indirect, are in most cases used on isolated measurements, from either free-falling, free-raising or lowered instruments.
It should be noted that the variance of the vertical isopycnal strain has also been been used as an alternative to microstructure shear measurements, in particular to determine vertical diffusivity \citep[see e.g.][]{garabato_widespread_2004,kunze_global_2006}.

Here, we employ an alternative method to estimate {$L_O$ and} $L_T$, and thus the turbulence dissipation rate, using temperature variance at fixed depth, measured by moored instruments.
The temperature variance is used to compute the Ellison scale, which we then relate to the Thorpe scale.
We define the Ellison scale as:
\begin{equation}
  L_E = \left(\frac{\mathrm{d}\overline{\theta}}{\mathrm{d}z}\right)^{-1} \overline{\theta'^2}^{1/2}
  \label{eq:ellison}
\end{equation}
with $\theta'$ the {potential} temperature fluctuation around the (time) mean value $\overline{\theta}$.
{Throughout the work we will generally omit to mention that we use potential temperature rather than in-situ temperature unless stated otherwise (the difference is however very small in our data).}
The overline in (\ref{eq:ellison}) is understood as time averaging, and will be discussed more in detail in section \ref{sec:Methods:Processing:Wavelet}.

We will show that the part of the temperature variance at frequencies immediately beyond the internal wave band is the one correlated with the turbulence dissipation rate.
This part of the spectrum is in fact the one most affected by fine structure contamination, i.e.~spectral contamination due to the advection of sharp gradients \cite{phillips_spectra_1971,gostiaux_fine-structure_2012}.
Comparisons between the Thorpe scale and the Ellison scale from laboratory experiments and numerical simulations are available in the literature, to our knowledge at least in \citet{itsweire_measurements_1984}, \citet{gargett_scaling_1988}, \citet{itsweire_turbulence_1993}, \citet{smyth_length_2000} and \citet{mater_relevance_2013}.
{\citet{moum_energy-containing_1996} compared Thorpe and Ellison scales using microstructure profiles from the ocean thermocline.
As here, they found good correlation between Thorpe and Ellison scales, but it should be noted that a different definition of the Ellison scale is used therein, based on fluctuations along the vertical direction rather than time fluctuations.
Possibly for this reason, they did not need to consider the distinction between internal waves and turbulent motions.}
Our aim here is to analyze in detail the covariance of these two quantities in an oceanographic context, focusing in particular on the role of time scales and on the potential for estimating turbulence dissipation rate.
From a practical perspective, the use of the Ellison scale can provide the advantage of being computed from local quantities, i.e.~temperature variance at a fixed depth and local stratification.
{We note, however, that the fact that $N$ appears in equation \eqref{eq:eps_LT} implies that the profile may have to be reordered in order to compute $N$ (depending on how $N$ is computed), as we actually do here.}

Fitting the classical Batchelor spectrum to temperature wavenumber spectra is a common approach for estimating temperature dissipation rate and turbulence dissipation rate \citep[see, e.g.,][for an oceanographic application]{klymak_oceanic_2007}.
{A} connection between the turbulence and the internal wave spectra is plausible, {as turbulence is believed to be generated in the ocean by internal wave breaking to a large extent.
However,} the actual link between the two is still unclear, as testified by the continuing efforts aimed at including strongly nonlinear effects in a theory of interacting waves \citep[][]{nazarenko_wave_2011}.
{This work may in fact be relevant also for this issue, even if it is not the focus of this work.}

Datasets obtained using NIOZ high sampling rate thermistors \citep[e.g.,][]{van_haren_detailed_2012} provide with a unique opportunity to study the temporal variability of turbulence in the ocean.
These thermistors, deployed in a mooring, can measure temperature continuously and independently for up to two years, with a temporal resolution of $1\,\mathrm{Hz}$ and a vertical resolution set by the thermistor separation along the mooring line, usually about $1\,\mathrm{m}$.
A growing set of measurements has been collected in various locations, ranging from tidally dominated shores \citep{van_haren_internal_2012} to seamounts \citep{van_haren_detailed_2012}, to the open and deep ocean \citep{van_haren_high-resolution_2009,van_haren_large_2011}.
In this paper, we will consider {two} data sets, which combine high temporal resolution with a reliable estimation of the vertical density profile.

{The} article is structured as follows: in section \ref{sec:Methods:Data} we briefly describe the data sets used.
In section \ref{sec:Methods:Processing} we describe the data analysis procedure.
Results are presented in section \ref{sec:Results} and discussed in section \ref{sec:Discussion}.
{Summary of the main points and conclusions follow} in section \ref{sec:Summary}.

\section{Methods}
\label{sec:Methods}

\subsection{Data}
\label{sec:Methods:Data}

The {two} data sets used in this work come from two different moorings, deployed from spring to early fall 2013 in the Atlantic Ocean on the slopes of Seamount Josephine (details of the moorings are given in table \ref{tab:moorings}).
Each of the moorings had more than one hundred ``NIOZ4'' thermistors \citep[an evolved version of the ones described in][]{van_haren_nioz3:_2009} taped on a nylon-coated steel cable, at intervals of $0.7\,\mathrm{m}$ (mooring 1) or $1.0\,\mathrm{m}$ (mooring 2).
The thermistors sampled temperature at a rate of $1\,\mathrm{Hz}$ with a precision higher than $1 \,\mathrm{mK}$.
The moorings were attached to a ballast weight at the bottom, and to an elliptical buoy at the top.
The high tension on the string, due to the high net buoyancy (approximately $400\,\mathrm{kg}$), guarantees that the string effectively behaves as a rigid rod, and that the mooring excursion both in the vertical and in the horizontal directions are small, as checked in similar previous moorings \citep{van_haren_detailed_2012}.
All the thermistors in mooring 1 performed satisfactorily for the whole deployment, with a noise level of approximately $5 \cdot 10^{-5}\,^\circ\mathrm{C}$.
On the other hand, in mooring 2, 7 thermistors out of 140 provide no data due to battery or electronic failure.
The data from the two different moorings {are} processed following the same procedure.
Calibration is applied to the raw data and the drift in the response of the thermistor electronics, visible over periods longer than a few weeks, is compensated for.
Overall, signal-to-noise ratio ($SNR$) in mooring 2 is smaller than in mooring 1, mainly due to the smaller (approximately 10 times) temperature variations at the greater depth of these moorings.
This limits the accuracy of our Thorpe scale estimates from mooring 2, as well as the accurate estimation of temperature variance.
We thus focus our analyses on mooring 1, and consider {mooring 2 only for the comparison and assessment} of the skill of the technique developed for mooring 1 in sub-optimal conditions.

\begin{figure}
   \noindent\includegraphics[width=0.49\textwidth]{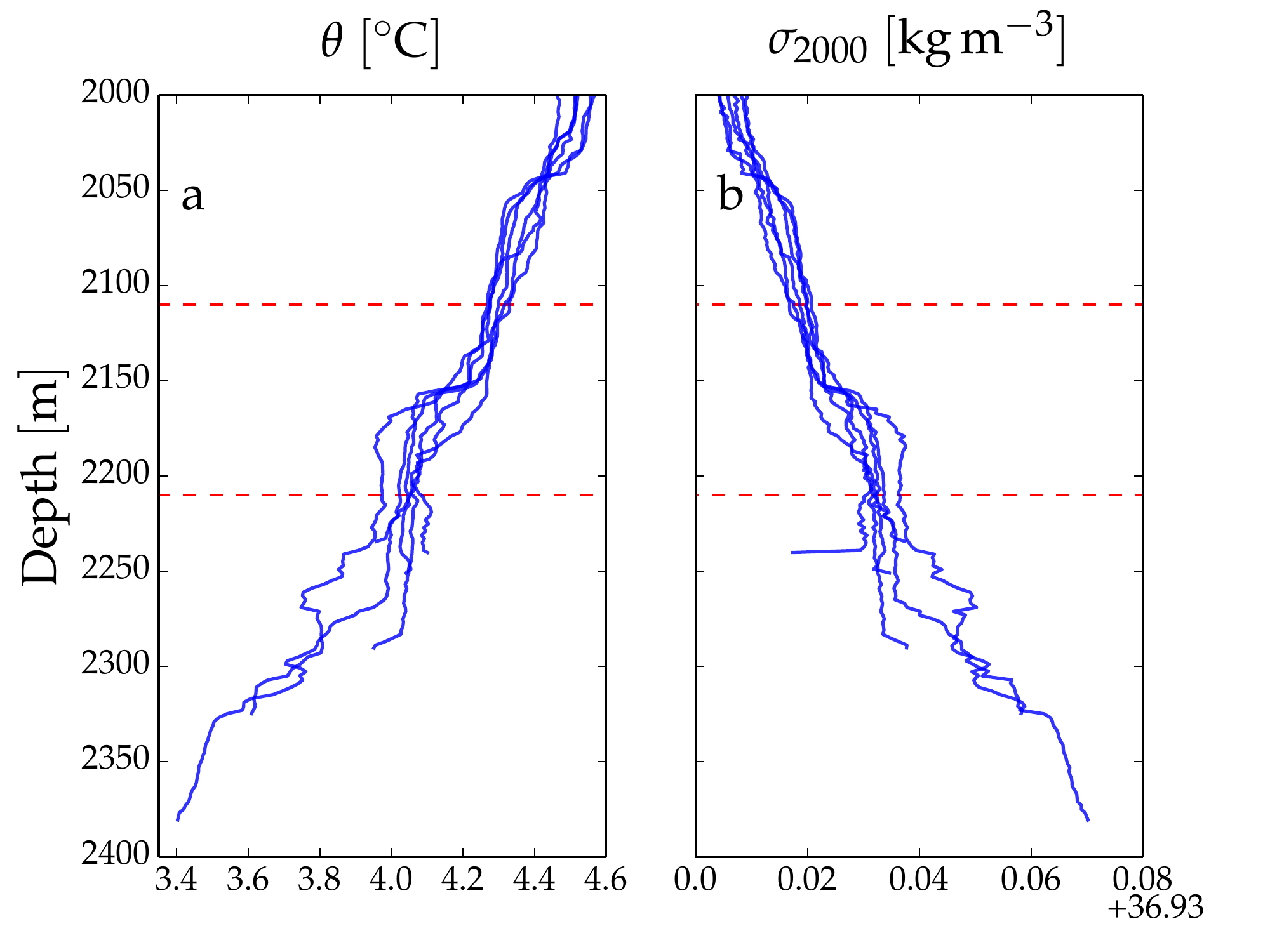}
   \noindent\includegraphics[width=0.49\textwidth]{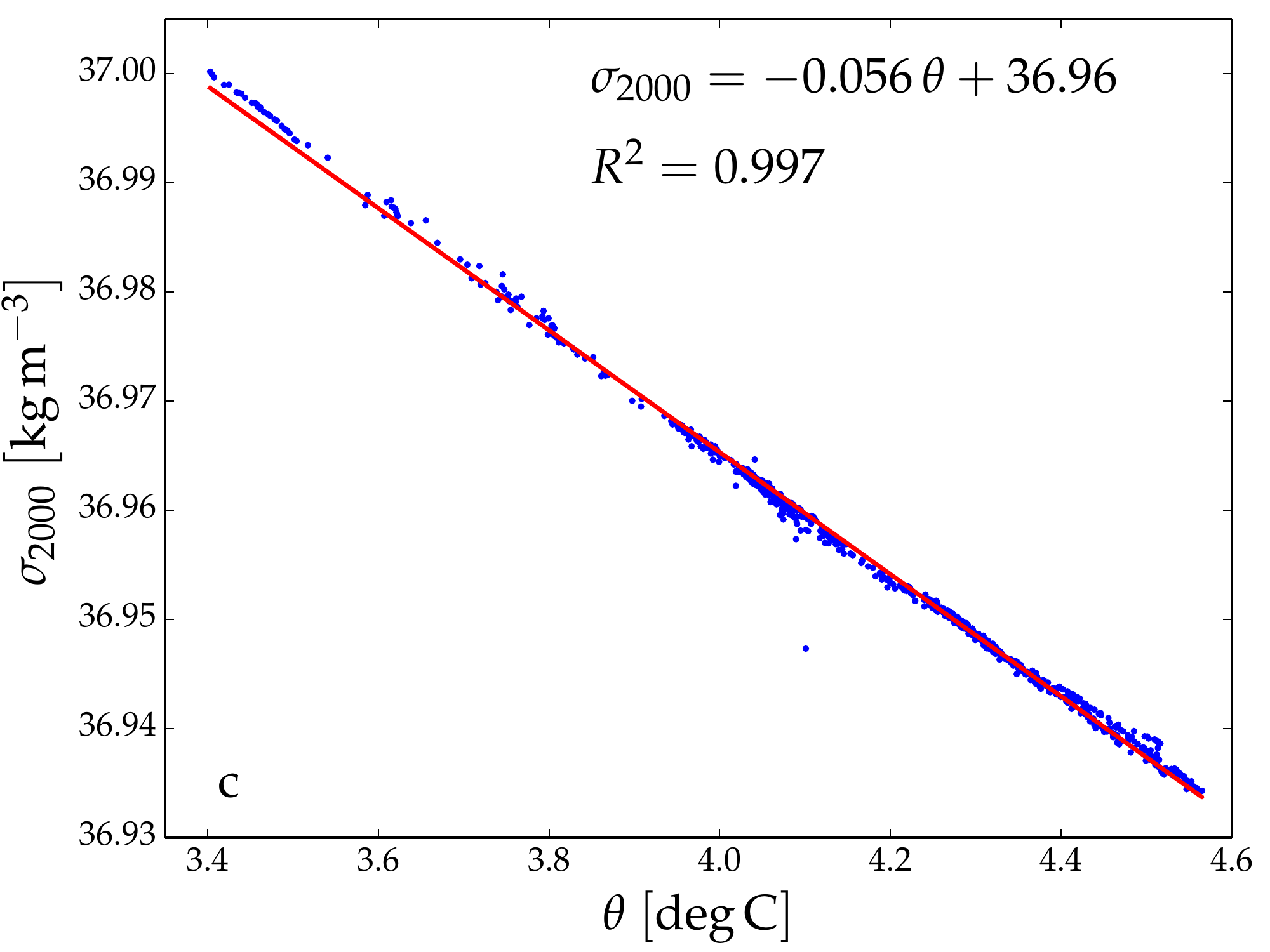}

   \noindent\includegraphics[width=0.49\textwidth]{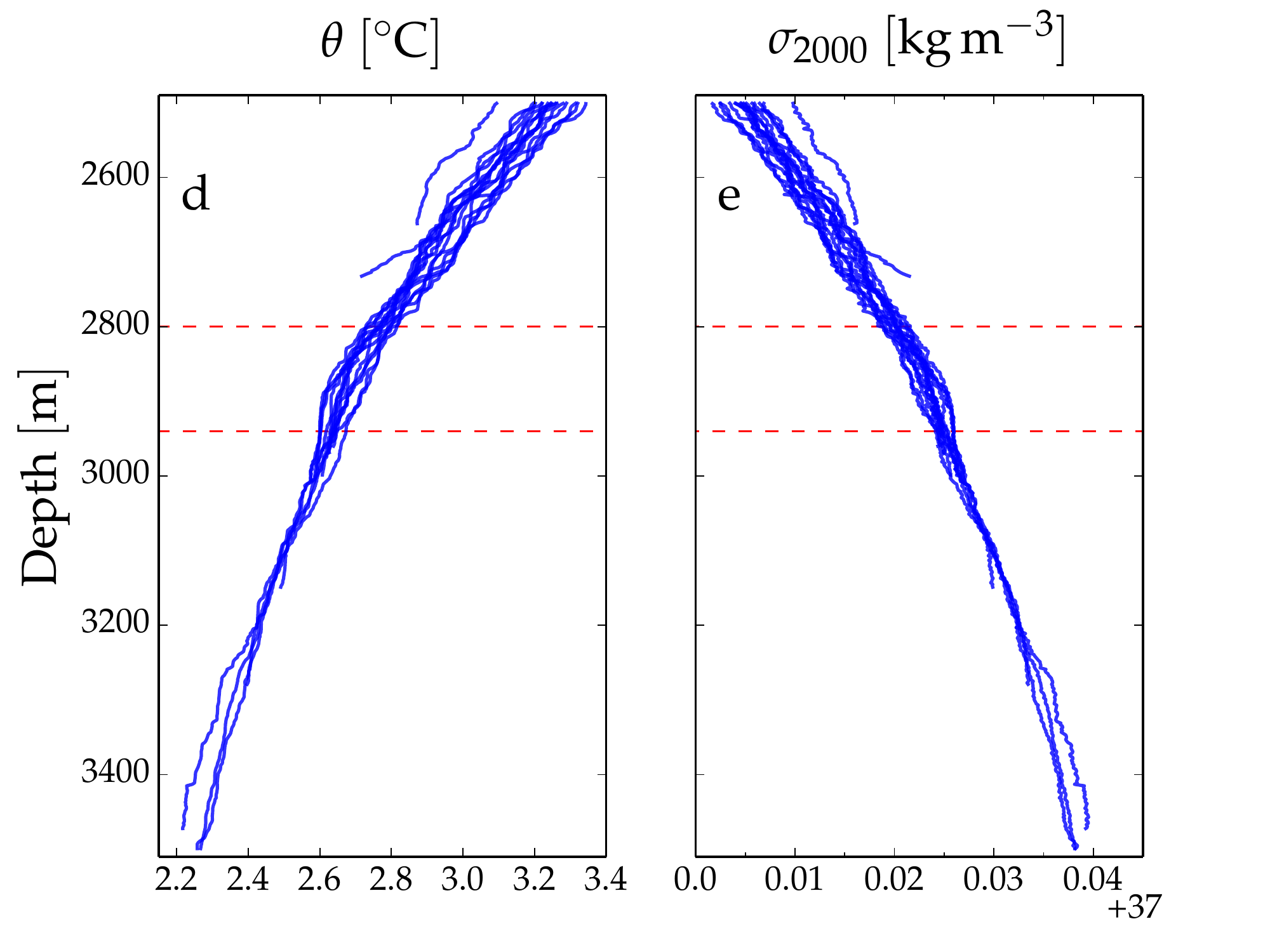}
   \noindent\includegraphics[width=0.49\textwidth]{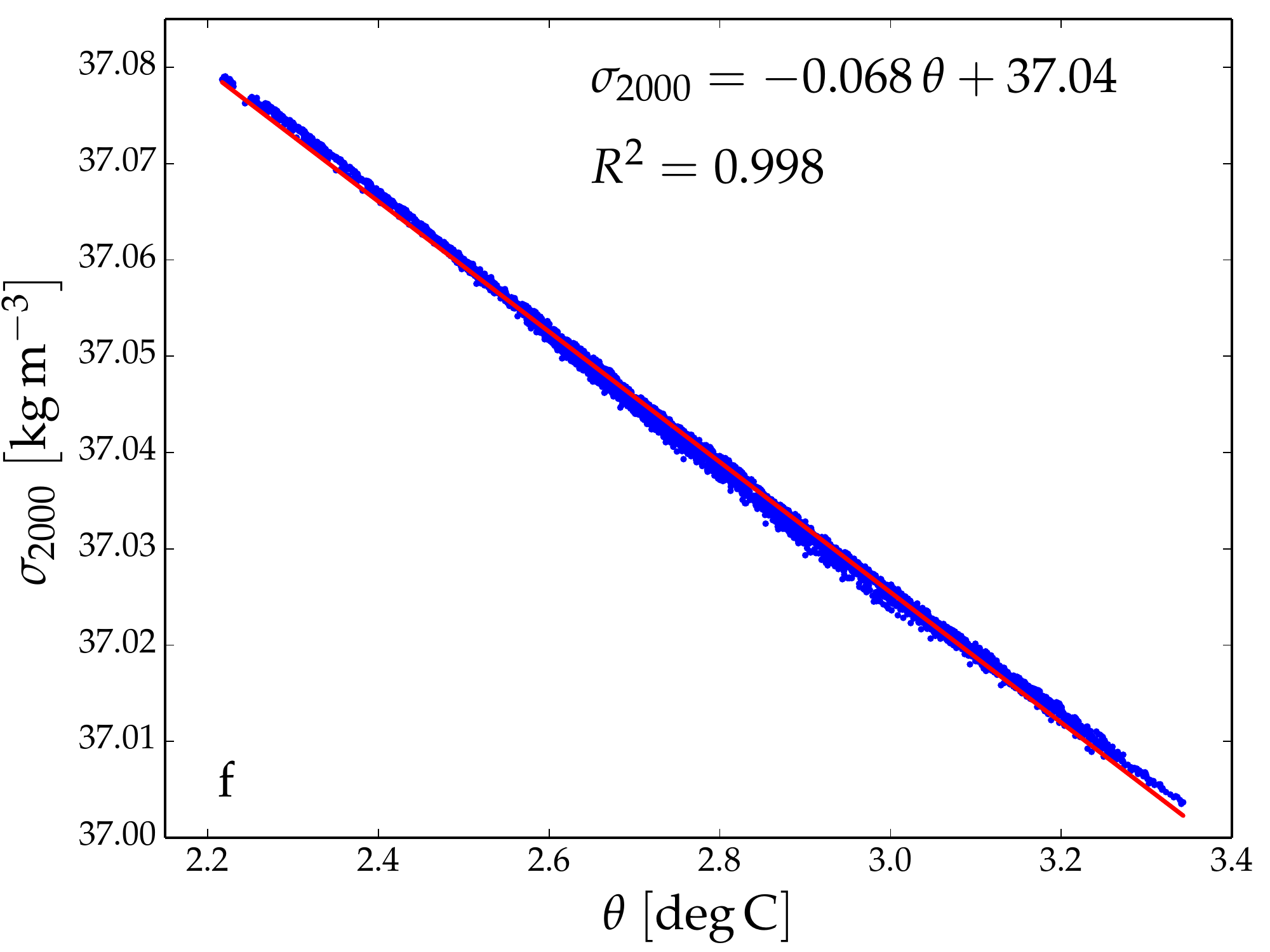}
    \caption{\label{fig:CTD}  Summary of the two CTD surveys performed.
      Panel a shows the potential temperature profiles and panel b shows the potential density anomaly profiles of the first survey.
      Similarly, panel d shows the potential temperature profiles and panel e shows the potential density anomaly profiles of the second survey.
      In panels a, b, d and e the dashed red lines mark the top and bottom depths of mooring 1 (panels a and b) and of {mooring 2} (panels d and e).
      The density--temperature relation in the first survey, for data {below $2000\,\mathrm{m}$}, is shown in {panel c}, {below $2500\,\mathrm{m}$} for the second survey in panel f.
      Data in all panels are $2\,\mathrm{m}$ bin averages from downward CTD casts.
      The results of a linear fit of the data are shown in panels c and f (see text).
      Note that the scales in the upper and lower panels are different.
      Noise is more prominent in the density profiles (panels b and e) than in temperature profiles (panels a and d) due to the combination of the conductivity and temperature measurements.
      Furthermore, panels b and e have a horizontal range which is smaller than panels a and d respectively, even if the thermal expansion coefficient is taken into account.
  }
\end{figure}

As mentioned in section \ref{sec:intro}, a key requirement for reliably estimating Thorpe scales and turbulence dissipation rate from temperature measurements is a tight {density--temperature} relationship.
On top of this, a linear density--temperature relationship in the measured temperature range further ensures that temperature is a good proxy of density.
In order to check if these assumptions are valid for the data considered here, two CTD surveys were performed.
The first one was performed after the recovery of mooring 1, sampling the water column approximately between $2000\,\mathrm{m}$ and $10\,\mathrm{m}$ above the bottom while the ship was moving at less than 1 knot towards the deployment location of mooring 2.
The results of this survey are summarized in figure \ref{fig:CTD}-a,b, which show $2\,\mathrm{m}$ bin averages of potential temperature and potential density anomaly respectively.
Figure \ref{fig:CTD}-c shows instead potential density anomaly as a function of temperature for $2\,\mathrm{m}$ binned data {below} $2000\,\mathrm{m}$.
The depth range is chosen in order to include approximately the same temperature range as the one recorded at mooring 1, {as well as measurements close to the bottom boundary layer} ($3.5-4.5\,^\circ\mathrm{C}$).
The data comes from the first six downward casts of the CTD, the ones closer to mooring 1.
Temperature from mooring 1 is higher by approximately $0.3\,^\circ\mathrm{C}$ during the last 20 days of the record than in the rest of the data set.
This suggests that warm water is being advected to the mooring location, possibly due to the passage of a vortex or a front.
As a consequence, this first CTD survey, performed during this warm phase, sampled warmer water than the thermistor string average at the same depth.
Furthermore, while moving away from mooring 1, slightly different water masses were found in the higher temperature range (close to $2000\,\mathrm{m}$).
This shows up in the bottom right part of figure \ref{fig:CTD}-c.
Despite this, density--temperature relationship is {well approximated by} a linear fit of the data, shown in the figure, {with} a coefficient of determination $R^2$ of $0.997$.
{A quadratic fit provides only a marginal improvement of $R^2$, and the coefficient of the quadratic term is approximately ten times smaller than that of the linear one.}

To better assess the linearity of the density--temperature relation, a more extensive survey was performed after the recovery of mooring 2.
The survey included 8 casts of the column {from $2500\,\mathrm{m}$ down to $10\,\mathrm{m}$ above the bottom} in a square region of side approximately $1\,\mathrm{km}$ centered at the mooring location.
Nine more casts over the whole water column {were performed}, along two perpendicular transects of length approximately $20\,\mathrm{km}$ centered at the mooring location.
The results of this second survey are collected in figure \ref{fig:CTD}-d, e and f.
Figure \ref{fig:CTD}-d,e show $2\,\mathrm{m}$ bin averages of potential temperature and potential density anomaly for the CTD casts of the second survey.
Figure \ref{fig:CTD}-f shows potential density anomaly as a function of temperature for $2\,\mathrm{m}$ binned data {below} $2500\,\mathrm{m}$ from all the downward CTD casts.
The data, spanning a broader temperature range than the one recorded at mooring 2, confirm the results of the first survey.
A linear fit of the data gives an $R^2$ of $0.998$, and we conclude that temperature can be considered a reliable proxy for density.
{Also in this case a quadratic fit does not give any substantial improvement over the linear one.}
The larger spread of the data in figure \ref{fig:CTD} {f} is partly due to considering all the casts together, and partly an artefact of the smaller temperature range plotted.
If taken separately, each profile has a tight linear relationship (not shown).
The potential density anomaly profiles in both figure \ref{fig:CTD}-b and figure \ref{fig:CTD}-e contain several overturns and {layers (the latter in particular in figure \ref{fig:CTD}-b), but the average stratification is approximately constant.}

\subsection{Data processing}
\label{sec:Methods:Processing}

\subsubsection{Thorpe scale analysis}
\label{sec:Methods:Processing:Thorpe}

The full data from the moorings, at a $1\,\mathrm{Hz}$ sampling rate, is low-pass filtered and subsampled to a time step of $25\,\mathrm{s}$ using non-overlapping moving averages, for reasons of computational efficiency.
The results are not sensitive to this subsampling, as was discussed first in \citet{van_haren_detailed_2012} and confirmed on subsets of the data here.
Missing data from mooring 2 (see section~\ref{sec:Methods:Data}) are linearly interpolated using data from the two thermistors directly above and below.
{In-situ temperature} is transformed to potential temperature by using a constant thermal expansion coefficient; the correction is, however, minimal.
The temperature profile at each time step is reordered to obtain the stably stratified reference profile, and to obtain an estimate of the Thorpe scale as the RMS displacement of water parcels in this process.
{To avoid extrapolation of stratification outside the mooring, fluid parcels are never displaced outside the top and bottom of the mooring.}
A threshold of $5\cdot 10^{-5}\,^\circ\mathrm{C}$ is used to remove overturns indistinguishable from noise.
When comparing these results to those from frequency spectra, further time averaging is performed, in order to match the lower time resolution of the $L_E$ time series (see section \ref{sec:Methods:Processing:Wavelet}).
For further discussion of the computation of Thorpe scales in a comparable data set, see \citet{van_haren_detailed_2012}.

\begin{figure}
   \noindent\includegraphics[width=0.49\textwidth]{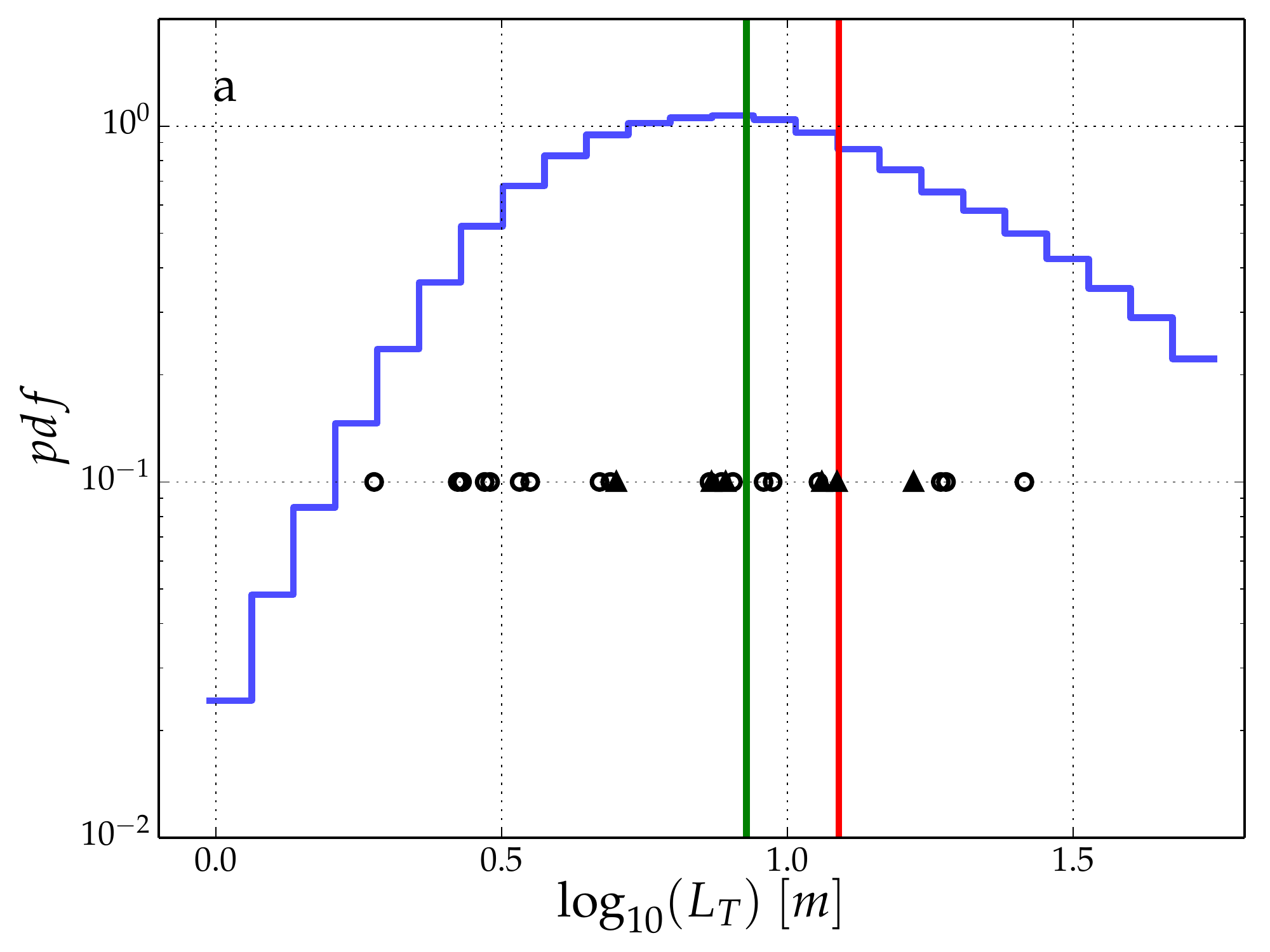}
   \noindent\includegraphics[width=0.49\textwidth]{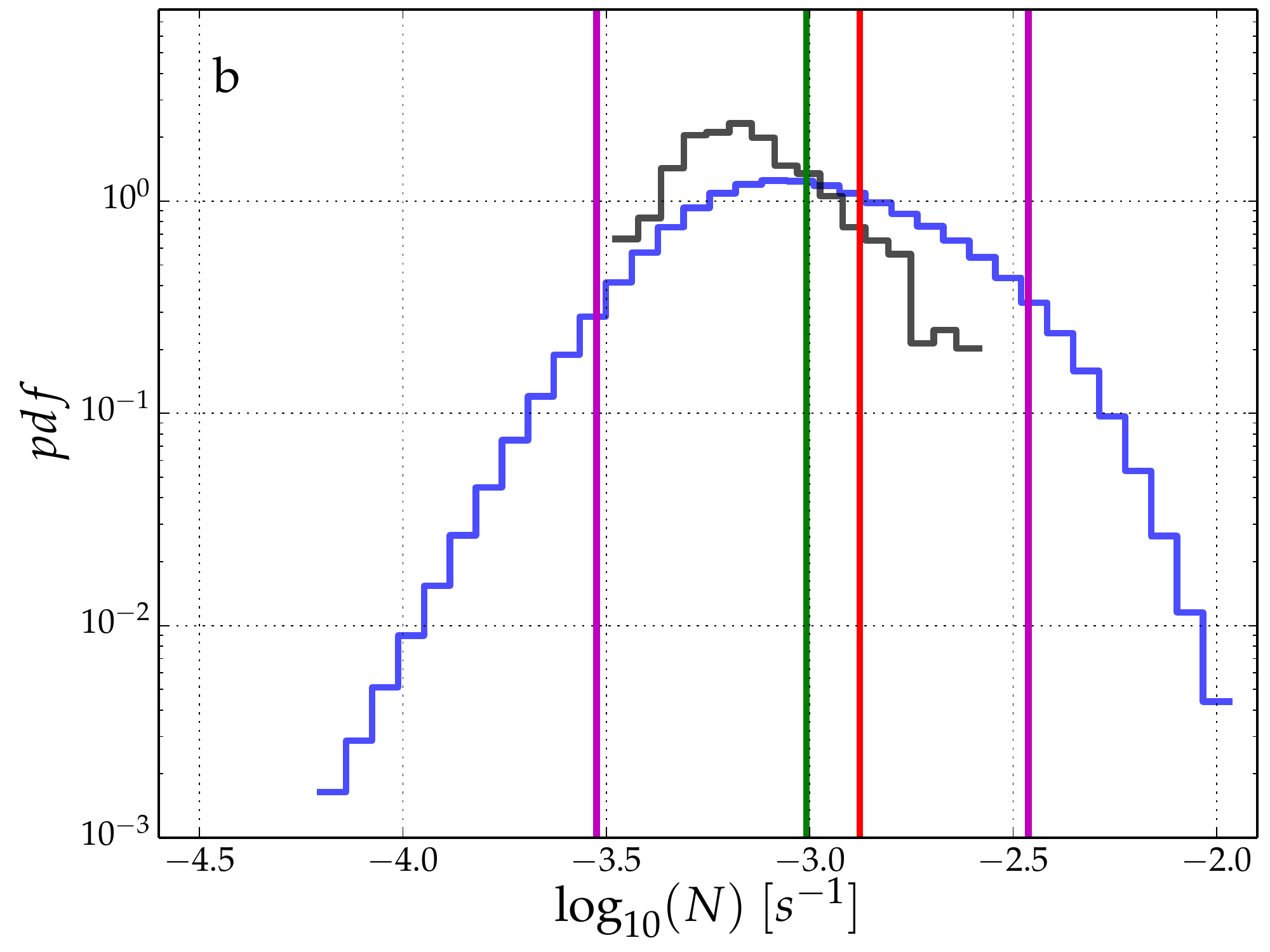}
   \noindent\includegraphics[width=0.49\textwidth]{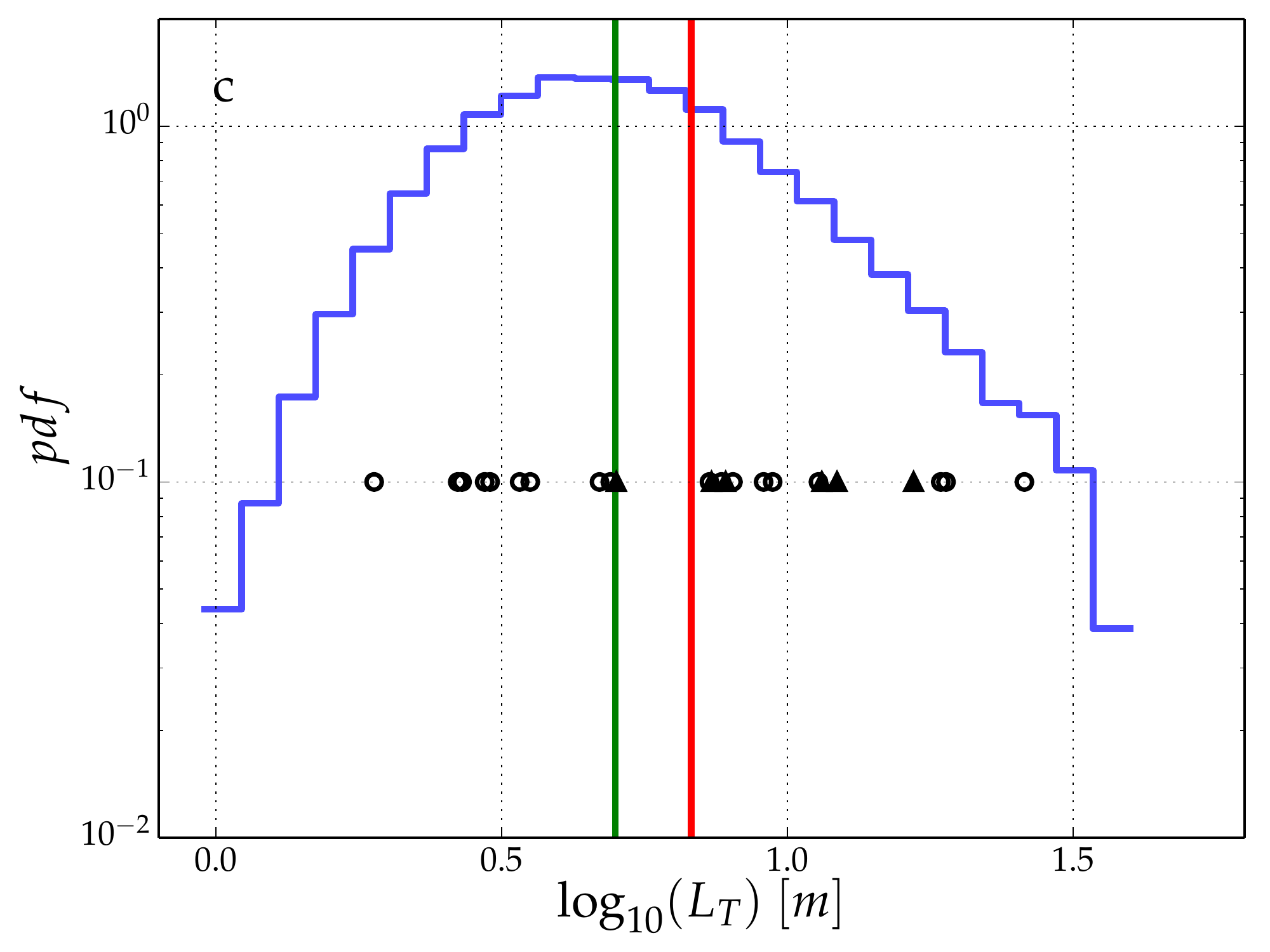}
   \noindent\includegraphics[width=0.49\textwidth]{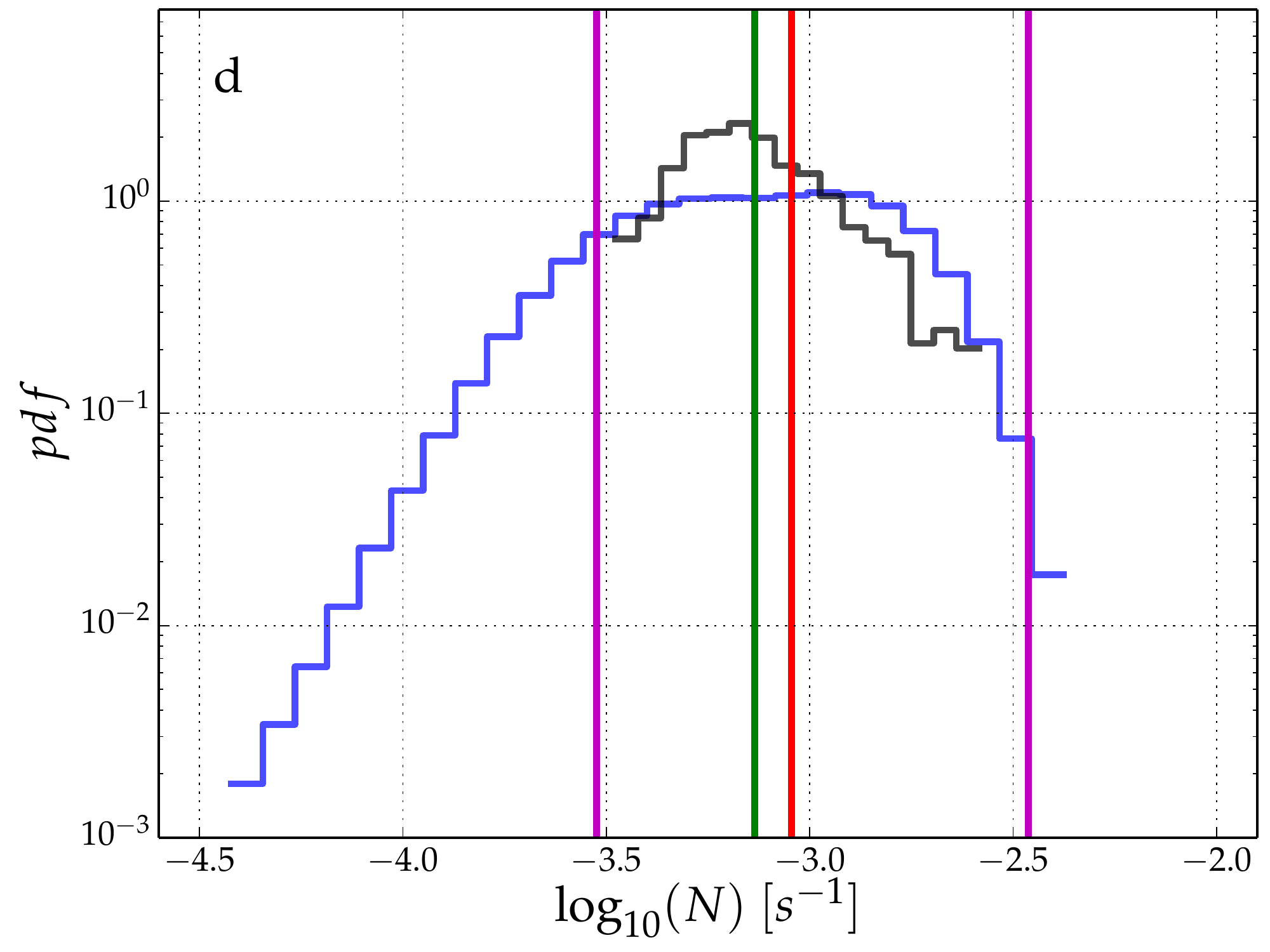}
    \caption{\label{fig:Thorpe} Probability density function of the logarithm of the Thorpe scale (mooring 1: panel a, mooring 2: panel c) and of the logarithm of the buoyancy frequency (mooring 1: panel b, mooring 2: panel d).
      The mean (vertical red line) and median (vertical green line) of the ensembles are also shown.
      In panels b and d, the pink lines mark $N_1$ and $N_{99}$.
      {RMS displacement from the CTD casts, computed within $100\,\mathrm{m}$ from the bottom in each CTD cast separately, is also shown with the black filled triangles (first survey) and the empty circles (second survey) in panels a and c.
      The probability density function of $2\,\mathrm{m}$-binned buoyancy frequency computed using the CTD data in the same depth range is shown in black in panels b and d.}
    }
\end{figure}

The probability density function of the logarithm of the Thorpe scales thus obtained is shown in figure \ref{fig:Thorpe}-a for mooring 1.
A typical size of the overturns is $10\,\mathrm{m}$, and we note that while the vertical resolution ($0.7\,\mathrm{m}$ for mooring 1) is sufficient to resolve the left tail of the distribution, the mooring length is imposing a cutoff to the right tail, i.e.~overturns larger than the mooring are present.
We thus have to be careful when comparing the largest values of $L_T$ to $L_E$, as only the former is limited by the mooring length.
{Note that the Thorpe scale is computed as the RMS of the displacement over one profile, and thus even the largest overturns, which locally give displacements of approximately $100\,\mathrm{m}$, are characterized by a Thorpe scale in figure \ref{fig:Thorpe} shorter than the mooring length (since we do not consider displacements longer than the mooring).}

The largest overturns, and those closer to the bottom, may also be influenced by the presence of the solid boundary, breaking the relation between Thorpe scale {and} Ozmidov scale, thus biasing the estimation of $\epsilon_{L_T}$.
{Since isolated overturns are not present in our dataset we cannot follow the classical approach of assigning one value of the Thorpe scale to each isolated overturn.
However, if we consider the local displacement (i.e.~the displacement needed to reorder the profile at each different depth), two clear patterns are visible.
The mean displacement is larger at the top and at the bottom, as the mean is dominated by the largest overturns, spanning the whole mooring.
If the RMS displacement is computed at each depth separately (i.e.~RMS of displacement at different times rather than at different depths), it is approximately constant throughout the mooring.
These two quantities seem to be affected by the solid boundary only in the region approximately within $5\,\mathrm{m}$ from the bottom of the mooring, with a decrease of both the mean and the RMS.
This gives us confidence that the boundary is not influencing the relation between Thorpe and Ozmidov scale strongly, in particular when column averages are taken.}
These issues are {even less} relevant at mooring 2 (figure \ref{fig:Thorpe}-c), for a combination of smaller overturns (typical size approximately $1/2$ the one at mooring 1) and the use of a longer mooring (see table \ref{tab:moorings}).

{The presence of the solid boundary may also affect the turbulent motions at small scales, breaking the relationship between Ozmidov scale and dissipation rate, and consequently invalidating equation \eqref{eq:eps_LT}.
We are however confident that this issue is not important here for various reasons.
First, the displacements are only marginally affected by the solid boundary, and only within the lowest part of the mooring.
Secondly, models of the dynamics above sloping boundaries with similar characteristics to those observed here indicate that the region where the classical ``law-of-the-wall'' model, with dissipation rate enhanced approaching the bottom, is relevant only within a thin layer $5$ to $10\,\mathrm{m}$ thick above the solid bottom \citep{slinn_modeling_2003,umlauf_diapycnal_2011}.
Previous measurements above sloping topography, using sensors to within $0.5\,\mathrm{m}$ from the bottom, showed that stratification virtually reached the bottom.
This suggests that this layer of intense turbulence close to the bottom may in fact be even thinner than what the models suggest.
}

{In order to check our computation of the Thorpe scale from the temperature profiles, we computed the same quantity from the density profiles of the CTD survey discussed in section \ref{sec:Methods:Data}.
The comparison of these two estimates cannot be very detailed, since the CTD and thermistor data only approximately overlap both in time and space, and also because the resolution and $SNR$ ratio of the CTD are lower than for the thermistors.
However, the RMS displacement computed within $100\,\mathrm{m}$ from the bottom (approximately the same range of the moorings used here) is consistent with the values from the thermistors, as shown in figure \ref{fig:Thorpe}-a,c.}

{Using} the reordered temperature profile, the buoyancy frequency can further be estimated at each time step, and $\epsilon_{L_T}$ can thus be computed using (\ref{eq:eps_LT}).
Both $L_T$ and $\epsilon_{L_T}$ are a function of time alone, since $L_T$ is the RMS of displacement of each vertical temperature profile, and the column average of $N$ is used (values at the top and bottom thermistors excluded).
In mooring 1, approximately $80\%$ of the points in the data set have non-zero displacement.
In mooring 2, approximately $50\%$ of the points have a non-zero displacement value.
Taking into account the fact that displacements over one overturn should sum up to zero, even higher overturning fractions are obtained in particular for mooring 1, for which the overturning fraction is close to $100\%$.

The probability density function of the buoyancy frequency is shown in figure \ref{fig:Thorpe}-b for mooring 1.
The mean values of $N$ are of order $10^{-3}\,\mathrm{s}^{-1}$ at mooring 1, first and $99$th percentiles of the distribution are $N_{1}\approx 2\cdot 10^{-4}\,\mathrm{s}^{-1}$ and $N_{99}\approx 5\cdot 10^{-3}\,\mathrm{s}^{-1}$.
At mooring 2 the distributions of $N$ are shifted to the left, with mean approximately $8\cdot 10^{-4}\,\mathrm{s}^{-1}$ and smaller first and $99$th percentiles ($N_{1}\approx 1\cdot 10^{-4}\,\mathrm{s}^{-1}$ and $N_{99}\approx 2\cdot 10^{-3}\,\mathrm{s}^{-1}$) than at mooring 1.
At mooring 2, $N$-distribution also has a broader peak than at mooring 1 (see figure \ref{fig:Thorpe}-d), but higher values of $N$ are markedly less frequent, with a very short right tail of the distribution.
All the statistics of $N$ are computed excluding the values {of} the top and bottom thermistors, since the value of $N$ is less reliable there, being computed using backward/forward differences instead of centered ones.
Figures \ref{fig:Thorpe}-b,d {suggest} that the vertical resolution of the two moorings imposes a cutoff to the highest values of $N$, in particular in mooring 2, having a vertical resolution of $1\,\mathrm{m}$, to be compared with the $0.7\,\mathrm{m}$ of mooring 1.
The presence of this cutoff is, however, less evident than for $L_T$ in figure \ref{fig:Thorpe}-a.
This cutoff is a consequence of resolving only a small portion of the temperature variance {turbulent} cascade, an unavoidable limitation of these kind of measurements, shared by all but the highest resolution temperature measurements in the ocean.
{To further check the results from the thermistors, $N$ is computed within $2\,\mathrm{m}$-high bins using the CTD data, in the region within $100\,\mathrm{m}$ from the bottom.
The probability density function of the buoyancy frequency thus computed is shown with a black line in figures \ref{fig:Thorpe}-b,d, and confirms that from the thermistors.}

\subsubsection{Ellison scale analysis}
\label{sec:Methods:Processing:Wavelet}

In order to compute the Ellison scale according to the definition (\ref{eq:ellison}), the temperature variance has to be estimated.
The Thorpe scale is here the RMS of displacements over the whole mooring, as computed at each time step, i.e.~it is a function of time.
To compare the two length scales, the Ellison scale has to be computed as a function of time as well, and thus temperature variance has to be estimated over moving windows of the time series.
On top of this, if the variance is estimated from temperature fluctuations at all time scales, the estimate of $L_E$ will be strongly biased by the presence of internal wave motions, as noted already by \citet{itsweire_evolution_1986}, and confirmed here (see section \ref{sec:Results}).

For these reasons, we use a time--frequency decomposition of variance, which provides an estimate of variance as a function of time, considering fluctuations with time scales up to a cut-off $\tau$.
One of the most commonly used time--frequency decompositions is the wavelet transform, which effectively is a series of filters producing an octave decomposition of a time series.
In particular, we use here the ``maximum overlap discrete wavelet transform'' (MODWT) following the methods described in detail in \citet{percival_wavelet_2006}.
This wavelet method is mostly equivalent to the more commonly used ``continuous wavelet transform'' \citep[see, e.g.,][for a practical introduction]{torrence_practical_1998}, but provides some computational advantages \citep{percival_wavelet_2006} over the continuous version, and was chosen for this reason.
{The description of the method is given in Appendix~\ref{sec:appA}, and here we briefly introduce only two quantities.
The MODWT produces an estimate of the variance of temperature at each time step and in the interval of time scales $[2^j\Delta t, 2^{j+1}\Delta t)$, with $\Delta t$ the time step of the time series and $j$ the level of the MODWT.
We write this variance estimate $\left<W^2\right>_{j,t}^{M}$, where $M$ indicates that $M$ estimates of variance are averaged, with a non--overlapping moving average in time, to increase statistical significance.
The main quantity that will be used is $\Sigma_{J,t}^{M}$, which is given by $\Sigma_{J,t}^{M} = \sum_{j=1}^{J}\left<W^2\right>_{j,t}^{M}$ and represents the variance of the temperature time series at time scales up to $\tau=2^{J+1}\Delta t$.}

\begin{figure}
   \noindent\includegraphics[width=\textwidth]{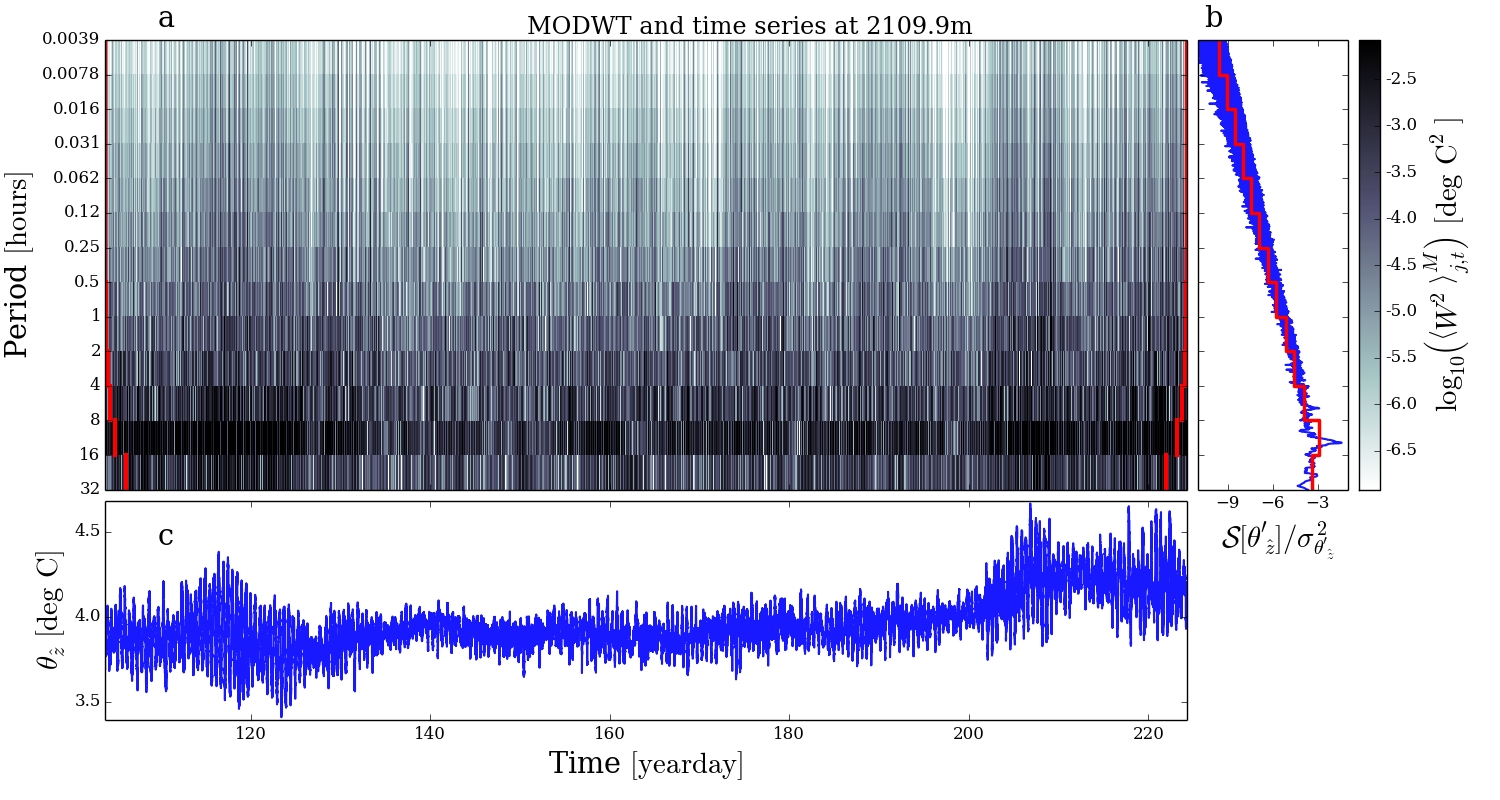}
    \caption{\label{fig:MODWT} Panel a: results for MODWT from a single thermistor at the top of mooring 1 (depth $2109.9\,\mathrm{m}$).
      The running mean of wavelet variance is shown, $\left<W^2\right>_{j,t}^{M}$, as a function of time and period.
      The results are from a ``least asymmetric'' wavelet filter of length 8, and use $M=101$.
      The vertical red lines mark the coefficient furthest away from each boundary which is influenced by boundary conditions (i.e.~computed using values outside the original time series, see text).
      The log color scale saturates at the $2$nd and $98$th percentiles.
      { Panel b: spectrum of the time series computed using a Fourier multi-taper method (blue line) and using the wavelet decomposition (red line).
      The spectrum is normalized by the time series variance.}
      Panel c: time series of temperature used in the MODWT shown in panel a and b.
      The time axis has units of yeardays, i.e.~1 January 2013 12:00UTC is yearday 0.5.
    }
\end{figure}

As an example, figure \ref{fig:MODWT}-a shows $\left<W^2\right>_{j,t}^{M}$ computed for the thermistor at the top of mooring 1.
Panel a of the figure shows the variance estimate for the time series in panel c as a function of time itself (horizontal axis) and time scale (vertical axis).
The results in the figure use $M=101$, the value that is used throughout the paper.
This value of $M$ implies that the time series of $\left<W^2\right>_{j,t}^{M}$ has a time step of $707\,\mathrm{s}$ (we are working on a temperature time series low-pass filtered and subsampled to a time step of $7\,\mathrm{s}$).
This value is a compromise between higher statistical significance (higher $M$) and time resolution (lower $M$).
Correlation between $\epsilon_{L_T}$ and $\epsilon_{L_E}$ (as well as $L_T$ and $L_E$) weakly depends on the value $M$, with higher $M$ giving slightly better correlation.
However, a relatively small value of $M$ is used in order to avoid reducing the observed range of values; averaging over longer time intervals removes the extreme values, and in particular leads to the undersampling of weakly turbulent phases.
Figure \ref{fig:MODWT}-b shows the spectrum of the same time series, computed both with a classical Fourier multi-taper method \citep[blue line, described for instance in the review of][]{ghil_advanced_2002} and by summing over all times the squared wavelet coefficients from the MODWT (red line, {see also appendix \ref{sec:appA}}).
The figure shows that most of the time series energy is concentrated in the low frequencies, which correspond to the semidiurnal tidal and to the inertial frequencies.
This gives further confidence that by low-pass filtering we neglect only a small part of the time series variance.
The frequency band containing the semidiurnal tide frequency (second from below in figure \ref{fig:MODWT}-a) is particularly prominent, indicating that the {temperature fluctuations} are dominated by {the tidal signal}.
It is also interesting to note that the energy is modulated at all frequencies, with particular strong variance at the beginning and end of the time series, most likely in connection with the passage of mesoscale or submesoscale features at the mooring location.
Slow modulations are highly correlated among different bands, suggesting that larger scale motions have a strong impact on the internal wave activity at all frequencies.

Once the MODWT is available, the computation of $L_E$ and $\epsilon_{L_E}$ is straightforward.
Using (\ref{eq:ellison}), $L_E$ is computed as:
\begin{equation}
  L_E = \frac{g \alpha}{N_\theta^2} \left(\Sigma_{J,t}^{M}\right)^\frac{1}{2},
  \label{eq:LE_wavelet}
\end{equation}
where we used the notation $\left.\frac{\mathrm{d}\overline{\theta}}{\mathrm{d}z}\right |_{\hat{z}} = -\frac{N^2_\theta}{g \alpha}$, with $g$ the acceleration of gravity and $\alpha$ the thermal expansion coefficient.
Equation (\ref{eq:LE_wavelet}) expresses $L_E$ as a function of the variance ($\Sigma_{J,t}^{M}$) in the time series up to the level $J$ of the MODWT.
The choice of the value of $J$ will be discussed in section \ref{sec:Results}.
{Note that $N_\theta$ in equation \eqref{eq:LE_wavelet} is just a shorthand for temperature stratification, and the physically relevant $N$, i.e.~the one in equation \eqref{eq:eps_LT}, may be different if temperature is not a good proxy of density.}

Similarly the $L_E$-based estimate of the turbulence dissipation rate $\epsilon_{L_E}$ follows from (\ref{eq:eps_LT}) and (\ref{eq:LE_wavelet}):
\begin{equation}
  \epsilon_{L_E} = 0.64\, \frac{(g \alpha)^2}{N_{\theta}^4}  \Sigma_{J,t}^{M} N^3.
  \label{eq:eps_wavelet}
\end{equation}
{The Thorpe scale} $L_T$ has been defined as a single value for the portion of the column measured by the thermistors (as the RMS displacement, see \ref{sec:intro} and \ref{sec:Methods:Processing:Thorpe}).
Consequently, the estimates obtained for each thermistor by using (\ref{eq:LE_wavelet}) and (\ref{eq:eps_wavelet}) will be averaged vertically to be compared with Thorpe scale based counterparts.
The importance of this averaging is further discussed in section \ref{sec:subsampling}.
{Also} in (\ref{eq:eps_wavelet}), we distinguish between $N_\theta$, the local temperature stratification which the thermistors always provide, and $N$, the physically relevant background buoyancy frequency, generally considered to be an average value over the overturn.
The two will in general be different also due to the effect of salinity.
In these cases, the background $N$ {and its variation in time} has to be recovered from other measurements, which will usually be profiles from shipborne CTD measurements.
In this work, we will not consider in detail this latter case However, {since $N$ is here well approximated by $N_\theta$, as discussed in section \ref{sec:Methods:Processing:Thorpe}.}.

\section{Results}
\label{sec:Results}

\subsection{Mooring 1}

\subsubsection{Time scales}

In the previous sections, the key point left unaddressed was the estimation of $\tau$, i.e. the cut-off time scale to exclude internal wave motions from the estimate of $L_E$.
This point can now be addressed by evaluating the validity of (\ref{eq:LE_wavelet}) and (\ref{eq:eps_wavelet}) applied to the dataset of mooring 1.
We evaluate the skill of our estimates by computing the cross-correlation between $L_T$ and $L_E$ (written in short $R[L_T,L_E]$), and between $\epsilon_{L_T}$ and $\epsilon_{L_E}$ (written in short $R[\epsilon_{L_T},\epsilon_{L_E}]$).
Table \ref{tab:correlation} collects the values of $R[L_T,L_E]$ and $R[\epsilon_{L_T},\epsilon_{L_E}]$ computed for different levels $J$.
We {recall the readers} that the value of $J$ determines the maximum period in the variance estimate $\Sigma_{J,t}^{M}$, entering equations (\ref{eq:LE_wavelet}) and (\ref{eq:eps_wavelet}).
Table \ref{tab:correlation} shows that the correlation starts to drop for $J$ greater than 7, i.e.~periods longer than $30\,\mathrm{min}$.
This period roughly corresponds to $2 \pi N_{99}^{-1}$ ($20\,\mathrm{min}$ for mooring 1).
Correlations lagged in time confirm that maximum correlation is attained for zero time lag (not shown).
Based on this evidence, we conclude that $\tau$ is approximately $20\,\mathrm{min}$.
If, however, we use $J=6$ or $J=7$ for our estimates, we find that $L_E$ systematically overestimates $L_T$ with the only exception of the largest overturns (see Appendix \ref{sec:appB}).
To avoid overestimation, we will use the estimates with $J=5$.

\begin{figure}
   \noindent\includegraphics[width=0.49\textwidth]{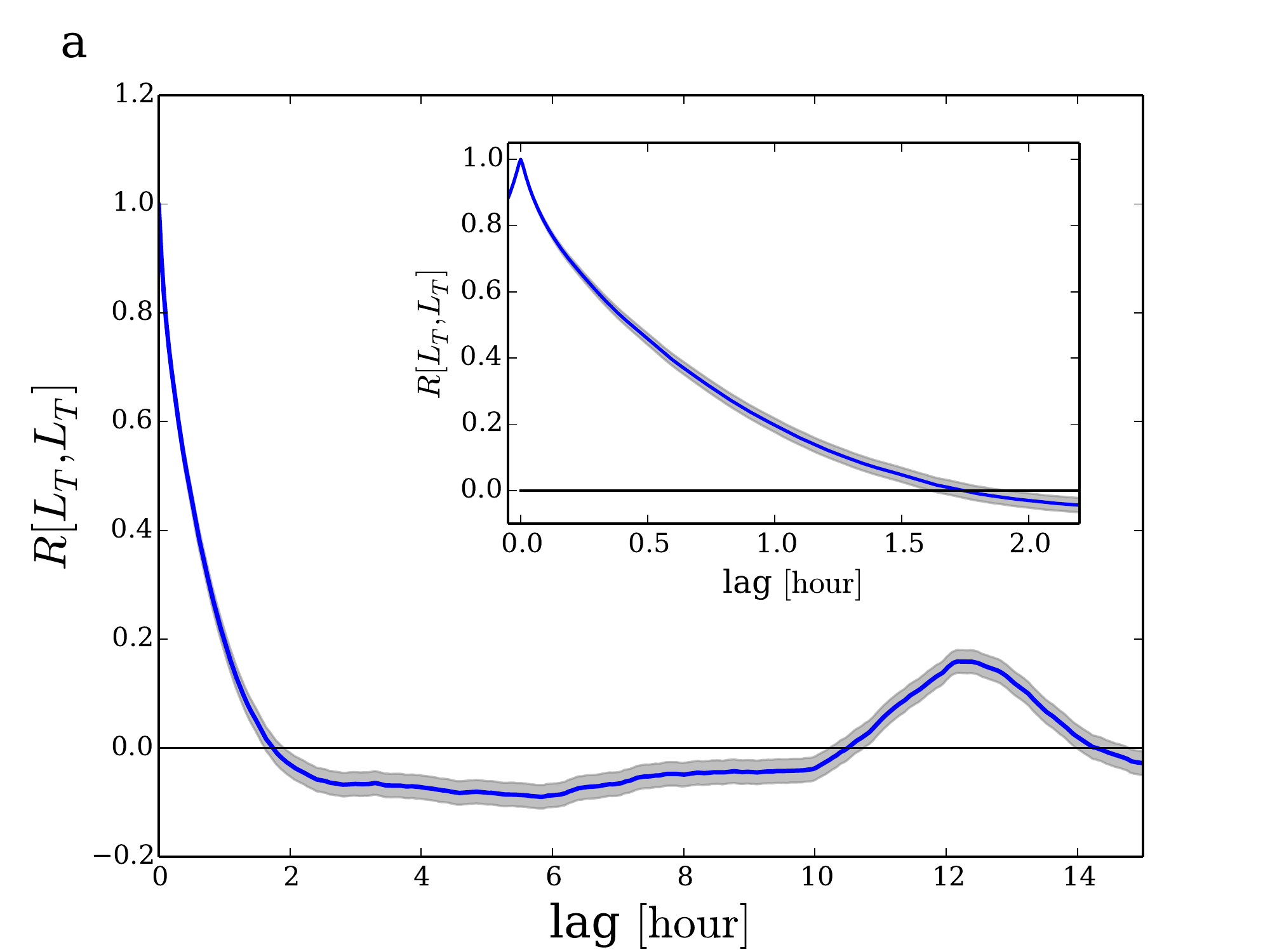}
   \noindent\includegraphics[width=0.49\textwidth]{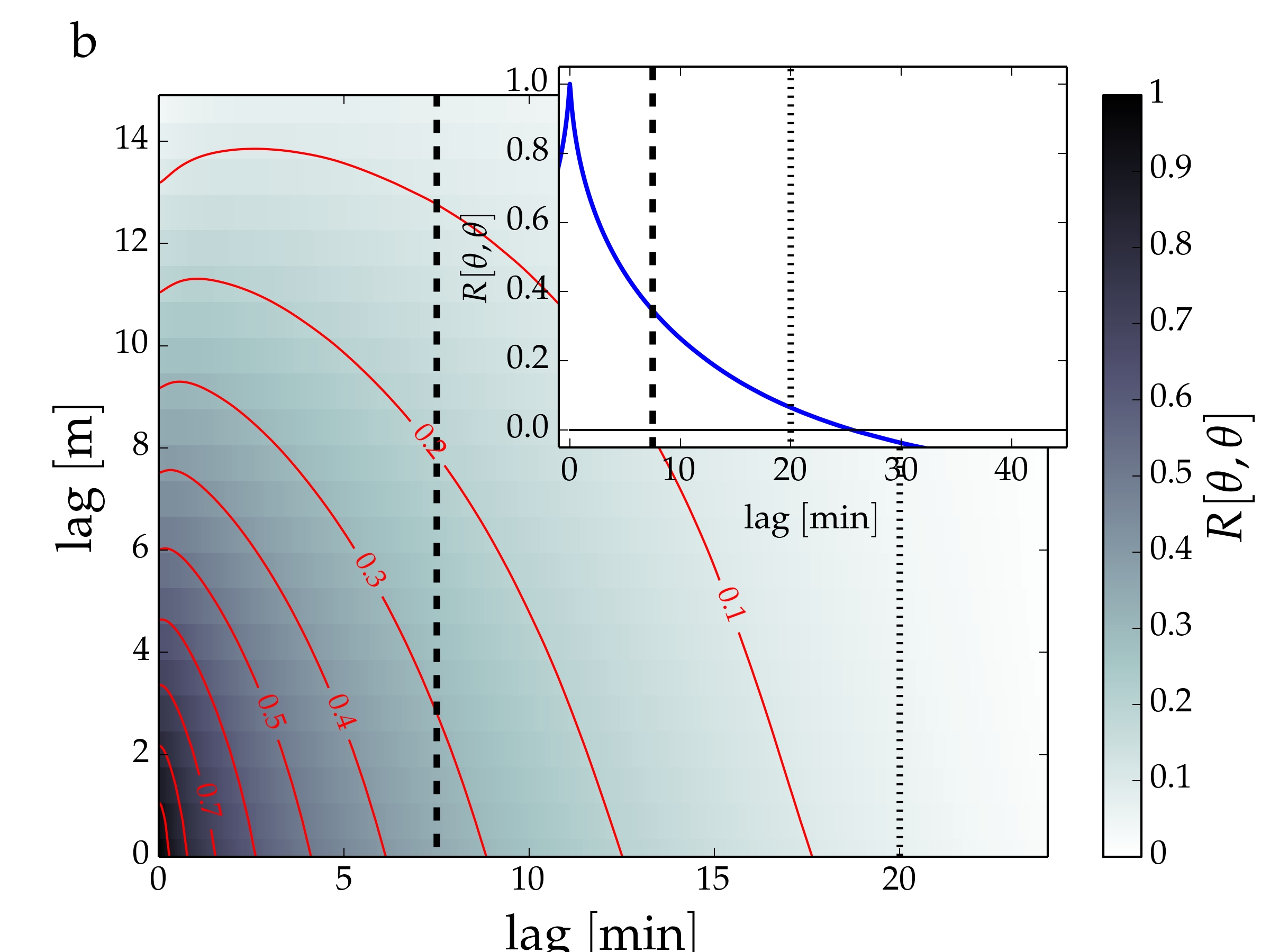}
    \caption{\label{fig:AutoCor}
      Autocorrelation function of Thorpe scale (panel a) and temperature (panel b) at mooring 1.
      Panel a shows correlation lagged in time, with the detail close to zero lag in the inset.
      The gray shaded area marks the 99\% confidence interval of the correlation.
      Panel b shows correlation lagged both in time (horizontal) and in depth (vertical), computed from the two dimensional, time--depth dependent, temperature field.
      In the inset of panel b, the autocorrelation for zero depth lag is reported.
      The red contours mark correlation values of $0.1$, $0.2$, $0.3$, $0.4$, $0.5$, $0.6$, $0.7$, $0.8$, $0.9$.
      The black vertical lines mark the maximum period nominally included in the estimates from MODWT (dashed) and $2\pi N^{-1}_{99}$ (dotted); see text for details.
    }
\end{figure}

It is worthwhile to analyze more in detail the process leading to maximum correlation for short time scales.
A water parcel is expected to return to its equilibrium position on a time of order $N^{-1}$.
{According to \citet{thorpe_turbulent_2005}, the typical decay time of Kelvin-Helmoltz instability may in particular be close to $30/N$.
Figure \ref{fig:AutoCor}-a shows the average correlation computed in linearly detrended subsets, lasting 4 days each, of the full time series of $L_T$ from mooring 1.
The inset in figure \ref{fig:AutoCor}-a suggests that an overturning patch typically lasts approximately 1.7 hours (autocorrelation crossing the zero line).
This value is consistent with the time scale expected for gravity wave motions with $N$ in the upper range of values found at mooring 1, but is shorter than the time scale expected from the mean $N$.
This suggests that horizontal motions are advecting the turbulent patches through the mooring.}
Time lags longer than one hour show anticorrelation, suggesting that an overturning event is more often followed by a quiescent period.
Figure \ref{fig:AutoCor}-a also shows that the semidiurnal tidal period of approximately $12.42\,\mathrm{hour}$ is dominant in the $L_T$ time series, confirming that the {turbulence} dynamics are to a large extent locked to the tide.

In figure \ref{fig:AutoCor}-b, we consider the temperature autocorrelation function.
Autocorrelation of temperature is computed here from the two dimensional, time and depth dependent, temperature record and is lagged both in time and in depth.
{A lag in depth means that the correlation is computed between copies of the data vertically shifted by multiples of the thermistor separation.}
The use of a two-dimensional correlation function has the aim of identifying the decorrelation time of temperature with itself, taking into account the fact that vertical displacements can reduce the correlation computed at a fixed depth.
{The time series is strongly self-correlated due to the tidal signal, at time scales longer than those of turbulence, and not relevant here.
For this reason, the computation is performed on subsets of the complete time series, each of length approximately $4.8\,\mathrm{hour}$.
Linear detrending of each subset is performed both in time and depth.}
The results in figure \ref{fig:AutoCor}-b show that this detrended temperature profile is not correlated with itself after approximately 30 minutes, and most of the correlation is lost already after 5 minutes.
Correlation is maximum for zero depth lag (correlation for zero depth lag is reported in the inset), most likely due to the fact that we are looking at time scales faster than most of the waves in the system.
The results from this brief study of the autocorrelation in the dataset thus support the conclusions drawn from table \ref{tab:correlation}.
The loss of self-correlation of the temperature signal is likely due to the incoherent (turbulent) motions above the internal wave band, as suggested by considering the time scale $2\pi N^{-1}_{99}$ (approximately $20\,\mathrm{min}$, drawn as a black dotted line in figure \ref{fig:AutoCor}-b), which corresponds to an autocorrelation at zero depth lag of approximately $0.1$.
{Loss of correlation is probably due to horizontal advection too.}
In practice, by using $J=5$ in the MODWT we limit $\tau$ nominally to $7.5\,\mathrm{min}$ (black dashed line), corresponding to an autocorrelation of temperature at zero depth lag of $0.4$.
We remember, however, that spectral leakage will lead, to some extent, to the inclusion also of longer periods.

\subsubsection{Length scales and turbulence dissipation rate}

\begin{figure}
   \noindent\includegraphics[width=0.49\textwidth]{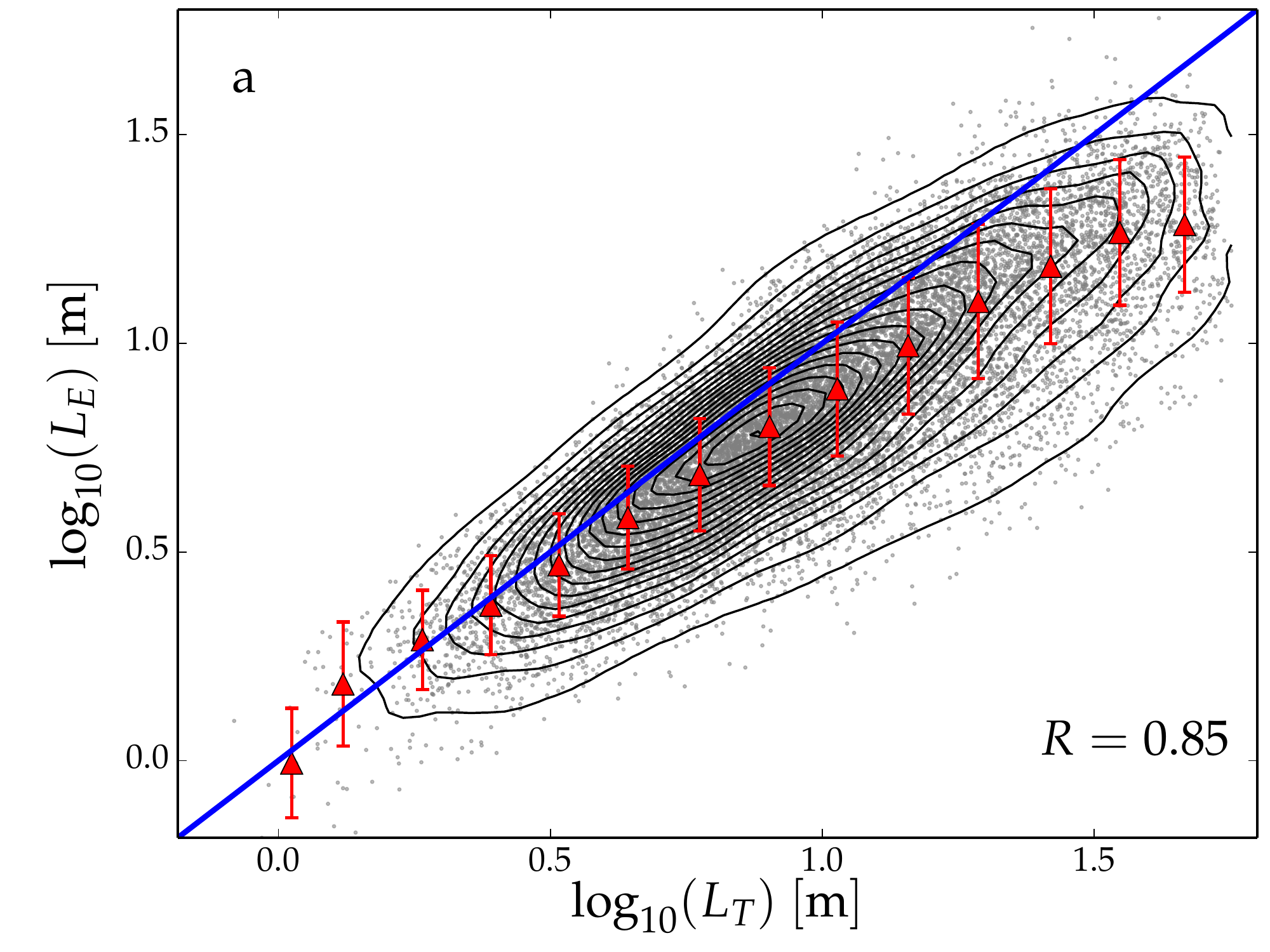}
   \noindent\includegraphics[width=0.49\textwidth]{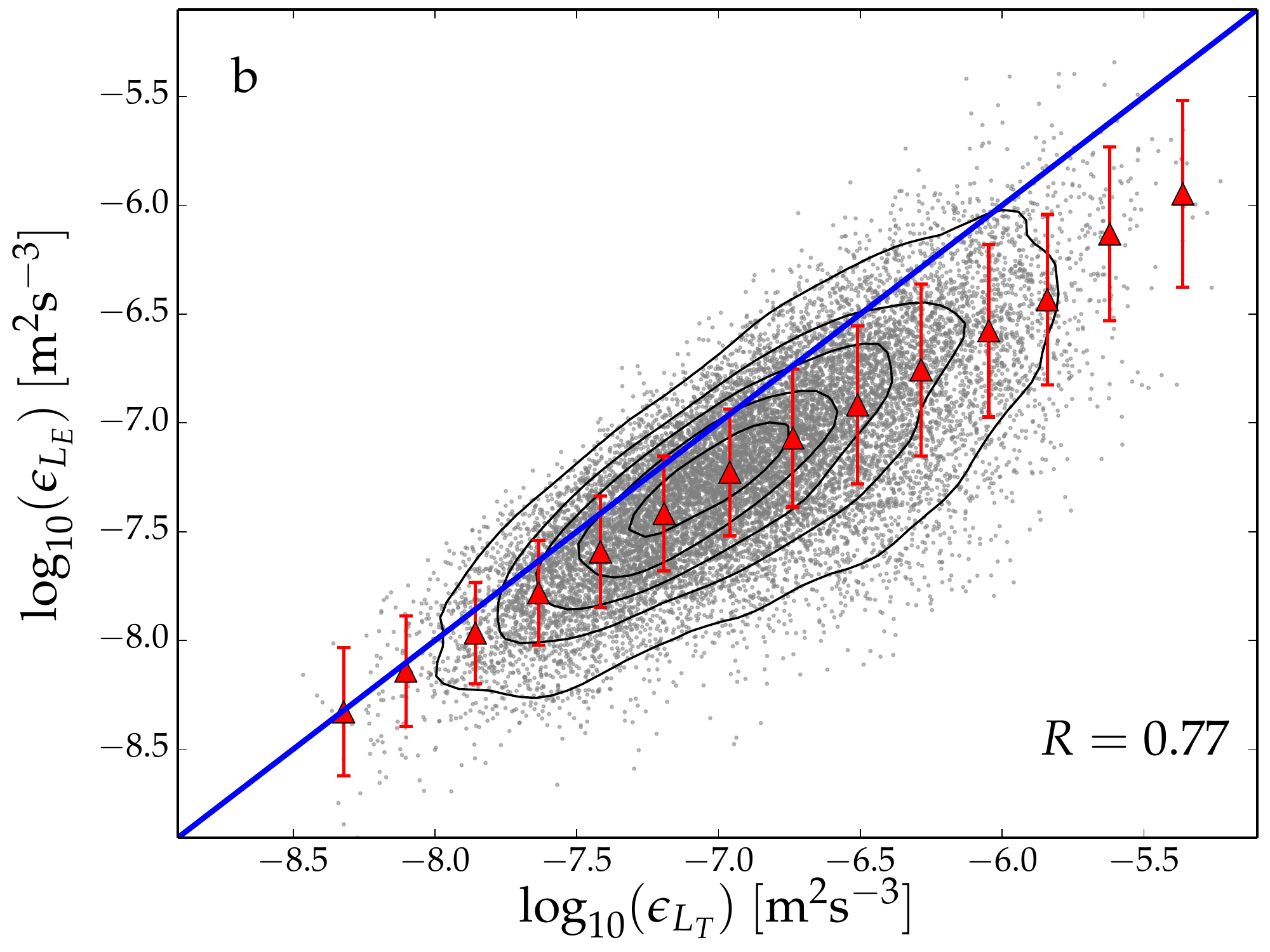}
  \caption{\label{fig:LTLE1}
      Results from mooring 1, using $J=5$ (longest nominal period $7.5\,\mathrm{min}$).
      Panel a: $L_E$ as a function of $L_T$.
      Each Gray dot is an estimate based on running averages over $707\,\mathrm{s}$ (see text).
      The black contours show the probability density function of the results, estimated with a Gaussian kernel \cite{jones_scipy:_2001}; the contours mark the values of the probability density function at intervals of $0.1$ starting from $0.1$.
      The blue line marks $L_E=L_T$.
      The red triangles mark the median of the distribution of points, computed in 14 bins; the error bars are the RMS of the distribution.
      Correlation between $L_T$ and $L_E$ is reported from table \ref{tab:correlation}.
      Panel b: same as panel a but for $\epsilon_{L_E}$ and $\epsilon_{L_T}$.
    }
\end{figure}

Figure \ref{fig:LTLE1}-a shows $L_E$ computed using equation (\ref{eq:LE_wavelet}) as a function of $L_T$.
In equation (\ref{eq:LE_wavelet}) we used $J=5$ and $M=101$, meaning that the quantities are all averaged to a time step of $707\,\mathrm{s}$.
We find that $L_E$ based on the MODWT provides on average a good estimate of $L_T$ computed by reordering the temperature profile.
Medians of the estimates are shown as red triangles in the figure, computed in 14 bins equally spaced in the $\log_{10}(L_T)$ space.
The dispersion of the estimates is less than a factor of 2 (0.3 in the logarithmic space of the plot), as measured by their RMS (red error bars in the figure).
$L_E$ underestimates, on average, $L_T$ at large values.
This has to be expected since (\ref{eq:LE_wavelet}) is based on a linear approximation, breaking down for larger values of the displacement.
On the other hand, there is a tendency to overestimation of $L_T$ by $L_E$ for smaller overturns, substantially less evident than in \citet{itsweire_measurements_1984} and \citet{smyth_length_2000}.
{This overestimation} is probably due to a combination of residual internal wave signal and to the non uniformity of the temperature gradient.
For a discussion of the effect of using different values of $J$ in the wavelet-based estimates, see Appendix \ref{sec:appB}.
Finally, we note that the length of mooring 1 imposes a cutoff for large $L_T$, already shown in figure \ref{fig:Thorpe}-a, which {apparently does not affect} the MODWT-based estimates of $L_E$.

Very similar conclusions can be drawn from figure \ref{fig:LTLE1}-b, which shows $\epsilon_{L_E}$ as a function of $\epsilon_{L_T}$.
Correlation is in this case slightly lower, but the method based on the MODWT still provides, on average, a good estimate of $\epsilon_{L_T}$.
The spread between the two quantities is also in this case larger for higher values, but the effect of the large scale cutoff in $L_T$ is not obvious as in figure \ref{fig:LTLE1}-a, since $N$ is included in the estimate (see equation \ref{eq:eps_wavelet}).
However, it is very likely that a significant number of the largest $\epsilon_{L_T}$ values is underestimated due to the cutoff in $L_T$.

\subsection{Mooring 2}

\begin{figure}
   \noindent\includegraphics[width=0.49\textwidth]{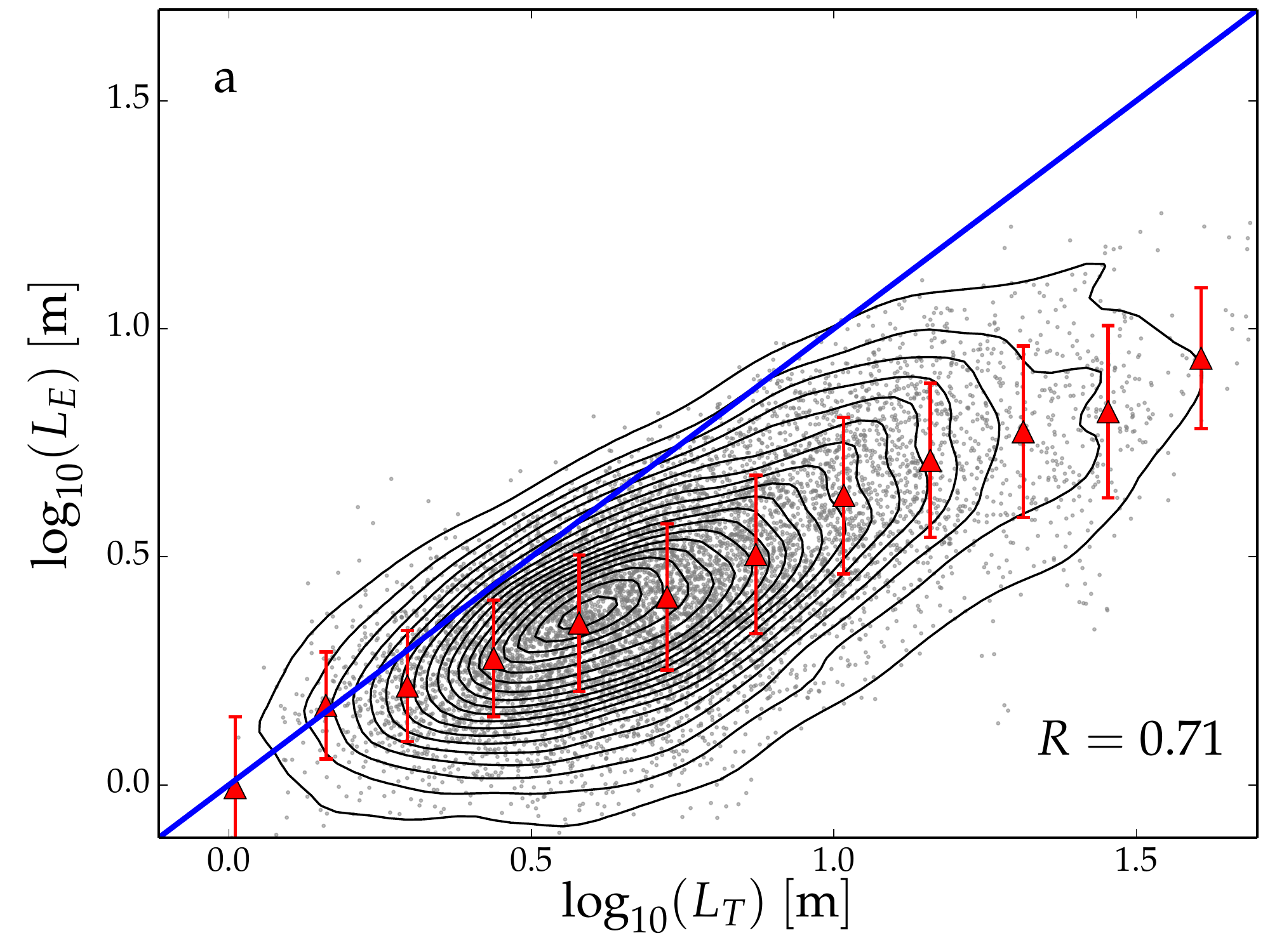}
   \noindent\includegraphics[width=0.49\textwidth]{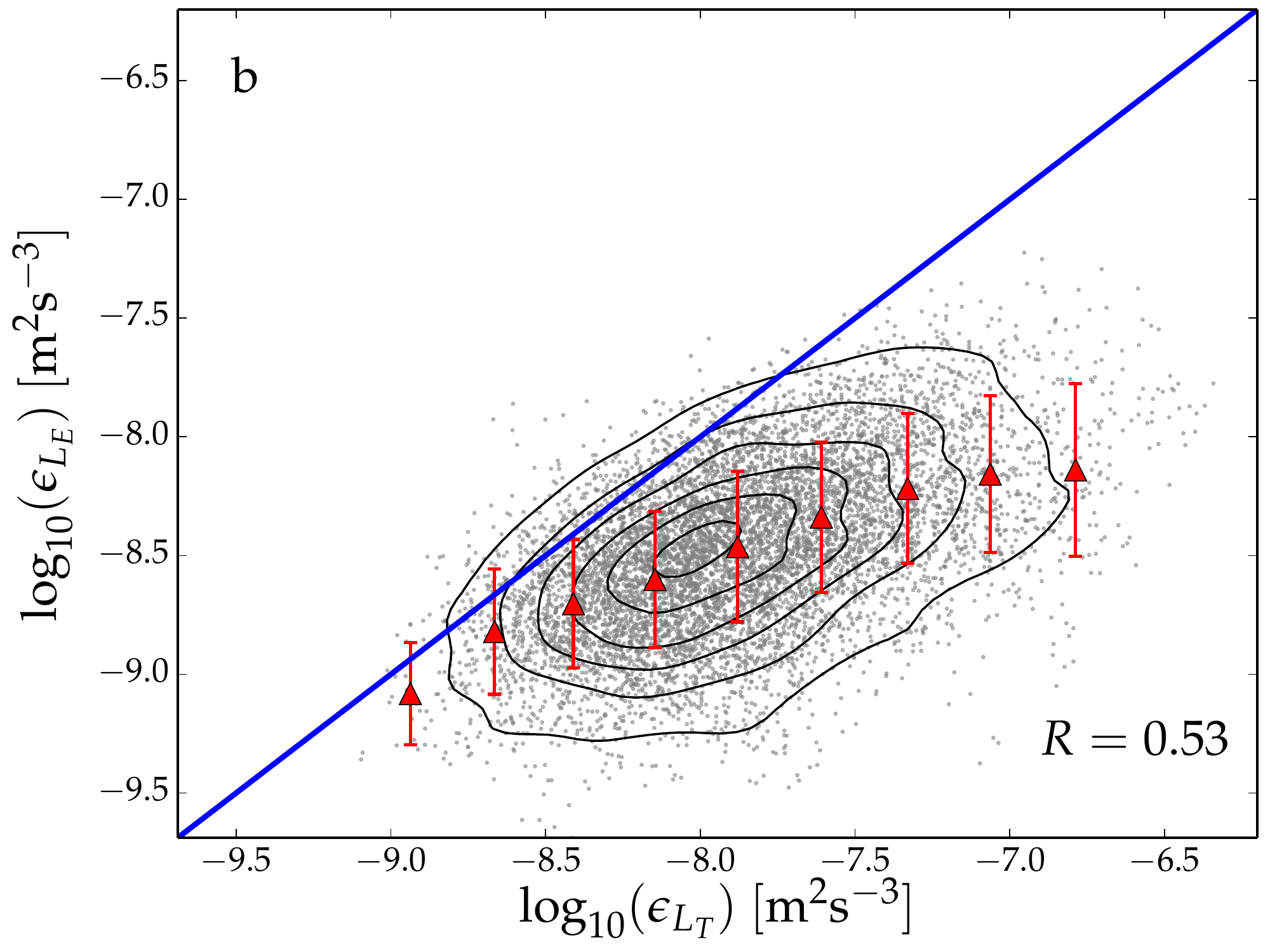}
  \caption{\label{fig:LTLE23}
    Same as figure \ref{fig:LTLE1} for mooring 2.
    For this mooring, $J=6$ was used (longest nominal period $15\,\mathrm{min}$, see text).
    }
\end{figure}

The analysis described for mooring 1 was applied to mooring 2, {the results being shown in figure \ref{fig:LTLE23}.}
In this case, the lower values of $N$ lead to the use of a higher $J$.
In particular, given that $2 \pi N^{-1}_{99}\approx 30\,\mathrm{min}$, we use $J=6$, i.e.~we include periods nominally up to $15\,\mathrm{min}$, {still providing high correlation as shown in table \ref{tab:correlation}}.
As discussed for mooring 1 this represents a conservative choice providing good results.
The same procedure applied to these two data sets leads to lower correlation between $L_T$ and $L_E$ and between $\epsilon_{L_T}$ and $\epsilon_{L_E}$ {in comparison with} mooring 1.
The reasons for the worse agreement are the broader distribution of $N$ (which makes the linear approximation worse) and the lower $SNR$ in these data sets as compared to the one from mooring 1.
Figure {\ref{fig:LTLE23}-a shows} that the largest $L_T$'s are well resolved, mostly due to the presence of few large overturns rather than due to the use of longer moorings.
This is clear also from figure {\ref{fig:LTLE23}-b, which shows} that the highest values of turbulence dissipation rate in {mooring 2 is} at least an order of magnitude smaller than in mooring 1.

Despite these differences, the results from mooring 1 are confirmed.
There is strong correlation between the quantities computed by reordering the temperature profile, and those computed by analyzing the variance in the temperature time series.
{Mooring 2 shows} that $L_E$ and $\epsilon_{L_E}$ systematically underestimate their Thorpe-scale-based counterparts for the largest events, a fact that was not clear from mooring 1 due to the cutoff of large overturns therein.
The Thorpe scale is in fact underestimated by $L_E$ also for smaller values in mooring 2.

Above $\epsilon_{L_T} \approx 10^{-7}\,\mathrm{m}^2\,\mathrm{s}^{-3}$, the $L_E$-based estimates {tend to underestimate more their $L_T$-based counterparts}, even if the high dispersion and the relatively few samples prevent a clearer identification of the change in behavior.
This underestimation is most likely a combined effect of the broader $N$ distribution and of the filtering procedure applied here.
In particular, by considering variance only in a portion of the frequency spectrum, we neglect the overturns associated with longer time scales (i.e.~the ones whose dynamics are linked to weaker stratification).
As discussed above, the $J$-level we use is a conservative choice, which can lead to underestimation of the Thorpe scales counterparts.
The variability of $N$ and the spectral leakage make different choices of $J$ possible, but a conservative approach has to be preferred in our view.
The filtering procedure is, in any case, essential to have a high correlation between the different estimates, as discussed with table \ref{tab:correlation}.
{We cannot rule out the possibility that the underestimation may in fact be the result of the overestimation of the Ozmidov scale by the Thorpe scale for the largest overturns.
This may be an important advantage of this technique, but a comparison with direct measurements of dissipation is necessary in order to understand if this is actually true.}

\subsection{Impact of spatial and time resolution}
\label{sec:subsampling}

\begin{figure}
   \noindent\includegraphics[width=0.49\textwidth]{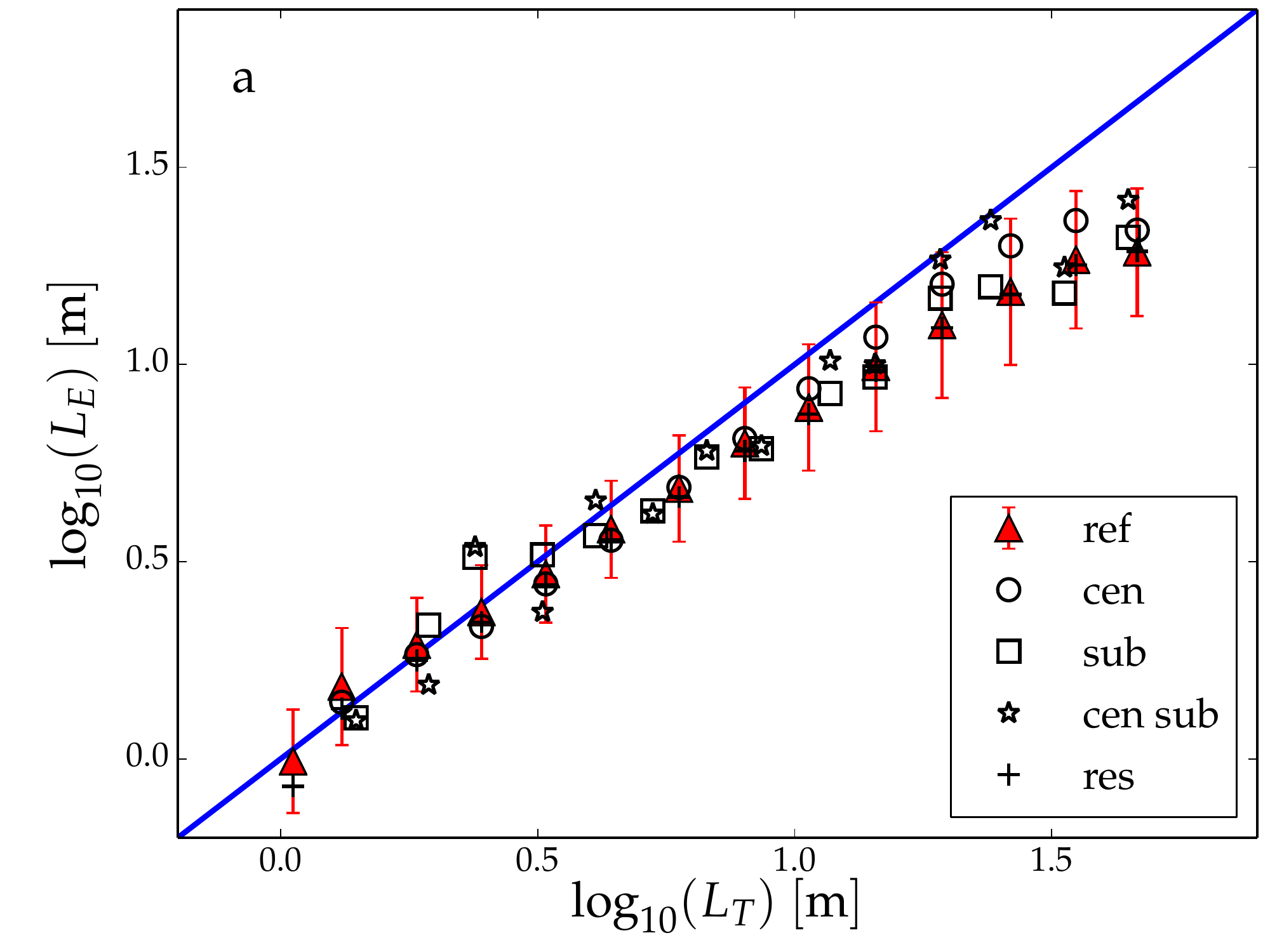}
   \noindent\includegraphics[width=0.49\textwidth]{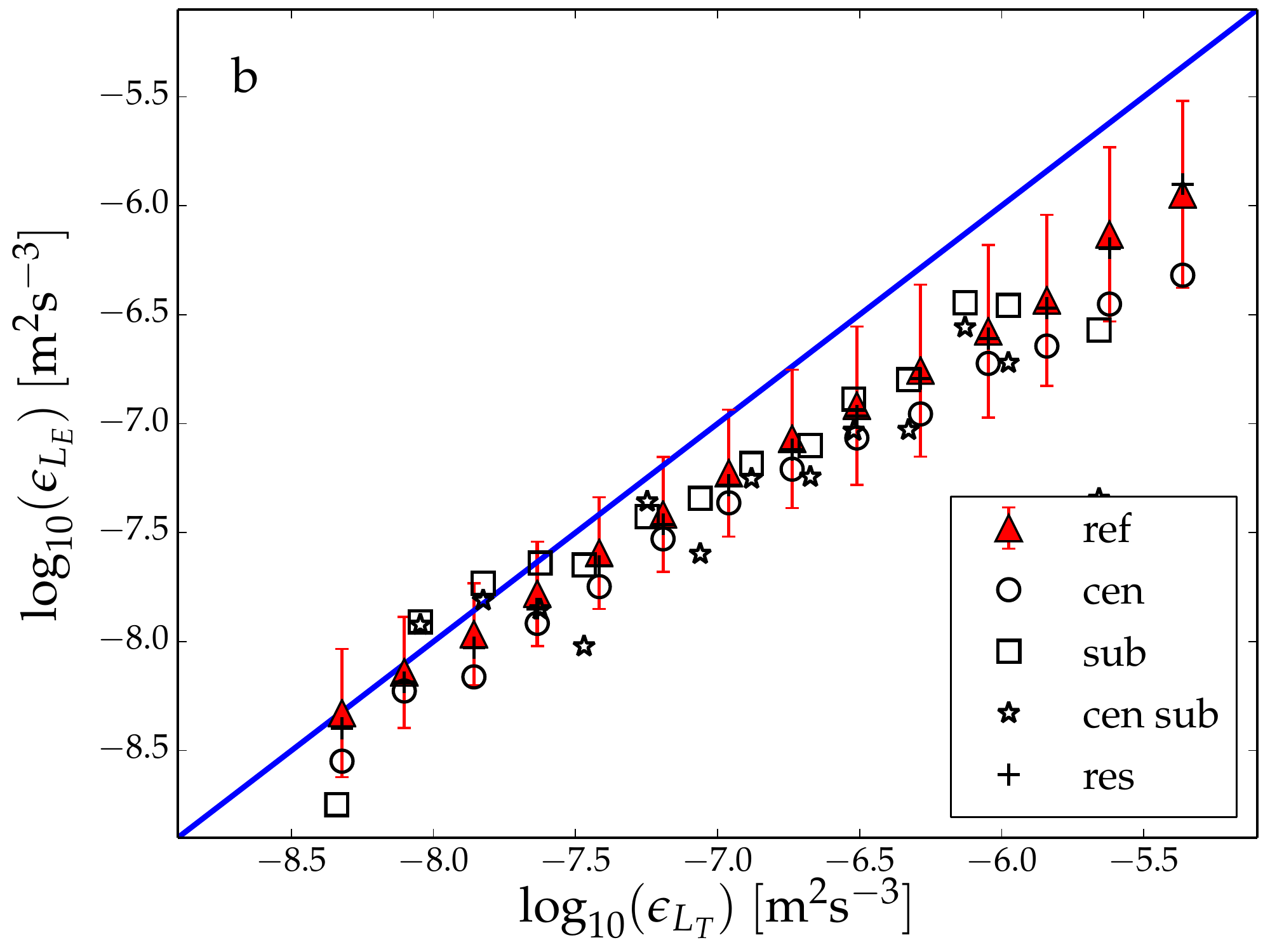}

   \noindent\includegraphics[width=0.49\textwidth]{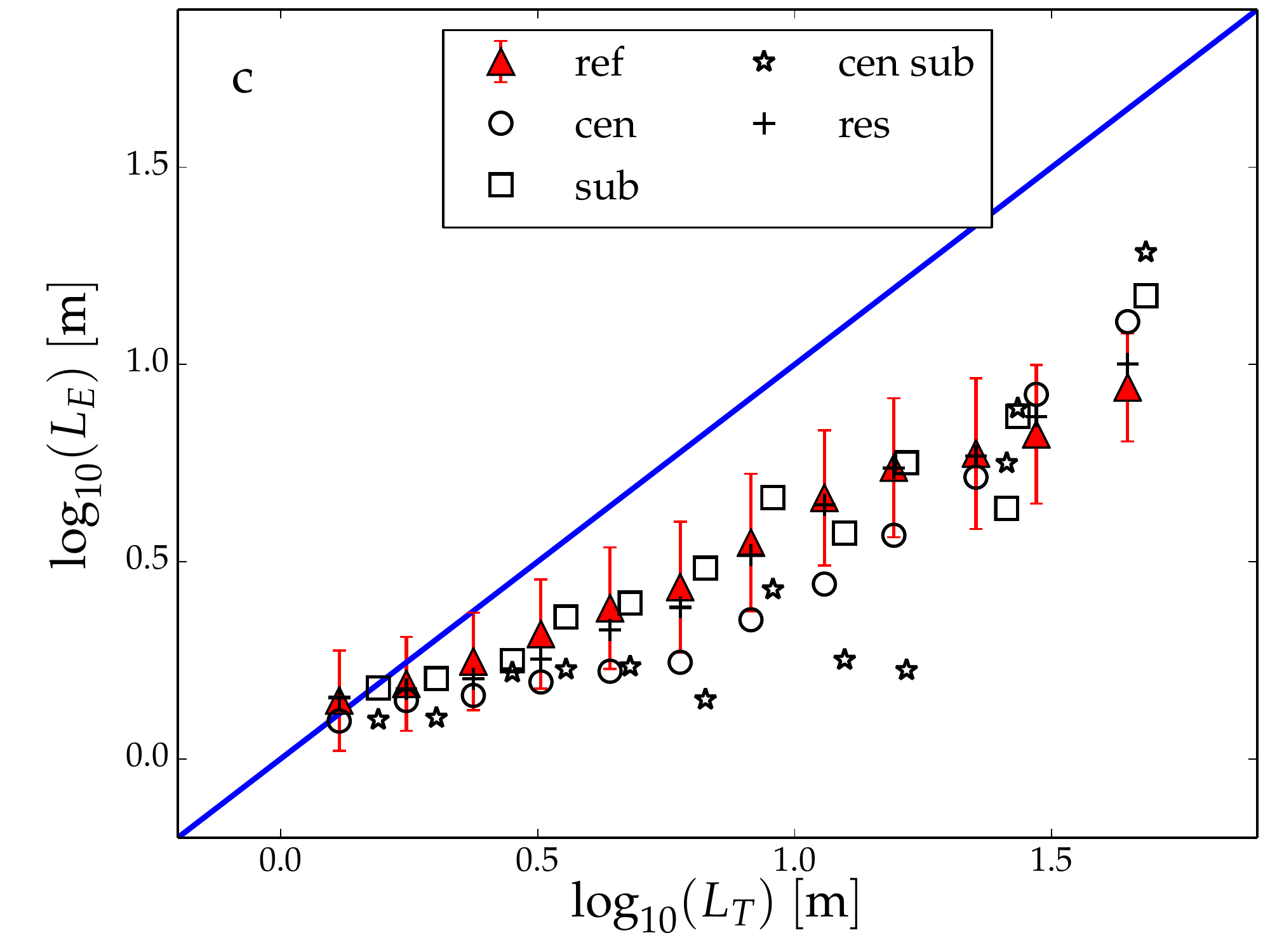}
   \noindent\includegraphics[width=0.49\textwidth]{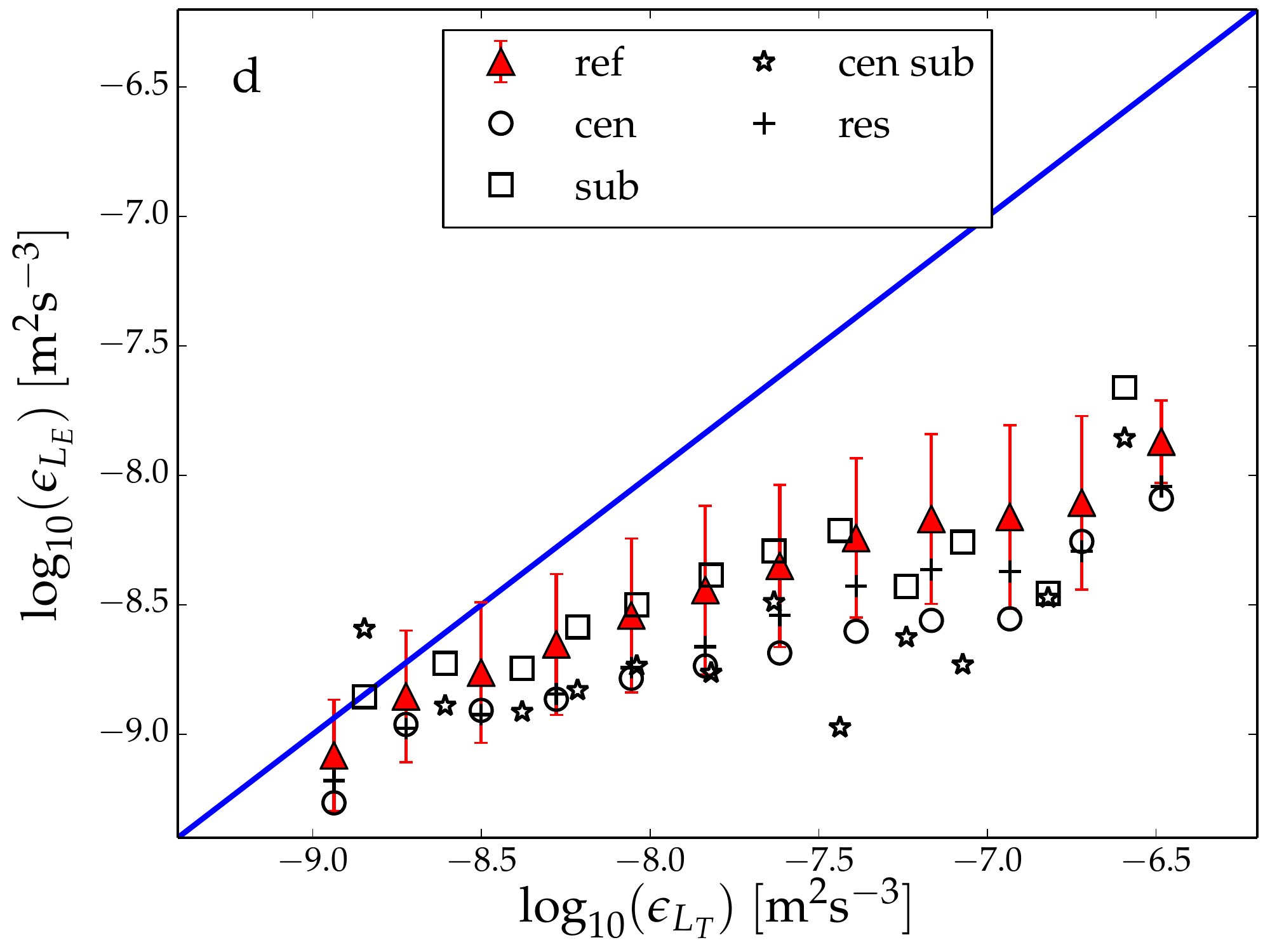}
    \caption{\label{fig:Subsampling}
      Similar to figure \ref{fig:LTLE1}, but showing only the median of the estimates for different kinds of subsampling (see table \ref{tab:subsamp} and text).
      Panels a and b show results from mooring 1, panels c and d show results from mooring 2.
      The results for $J=5$, shown in greater detail in figure \ref{fig:LTLE1}, are in red and with an error bar equal to the RMS of the distribution.
    }
\end{figure}

An interesting question to be asked is how sensitive the method is to changes in the spatial and time resolution.
In particular, we consider four cases in figure \ref{fig:Subsampling} (see also table \ref{tab:subsamp}), for both mooring 1 and 2.

First, we compute the MODWT estimates using only the thermistors occupying the central $1/5$ part of the mooring (case ``cen'', circles in figure \ref{fig:Subsampling}).
This estimate closely matches the one from the complete data set for mooring 1, in fact giving a better agreement with the Thorpe scale for the largest overturns (figure \ref{fig:Subsampling}-a, {circles systematically closer to the blue line than the triangles}), but a systematic underestimation of the turbulence dissipation rate (figure \ref{fig:Subsampling}-b).
For mooring 2, on the other hand, both overturn scales and turbulence dissipation rate are underestimated with respect to the use of the full data set.

Second, we assess the importance of time resolution.
The impact of a lower sampling rate in the original time series is discussed in Appendix \ref{sec:appC}.
Here, we consider instead the case in which temperature can be measured at the sampling rate used throughout this work, namely $7\,\mathrm{s}$ (see section \ref{sec:Methods:Processing:Wavelet}), but only for a limited amount of time every day.
In particular, we show in figure \ref{fig:Subsampling} the case in which the temperature variance is estimated twice a day over $707\,\mathrm{s}$ (the averaging period used in the MODWT analysis, see section \ref{sec:Methods:Processing:Wavelet}).
In the figure, these results are marked with a square (label ``sub'').
We see that for this case the agreement with the full data set is good, apart for the highest values of $L_E$ and $\epsilon_{L_E}$, having lower statistics.

The essential role of averaging is confirmed by computing the estimate on the central $1/5$ of the mooring, also using only two estimates per day (``cen sub'' case, stars in figure \ref{fig:Subsampling}).
The means of the distribution, shown in the figure, are noticeably noisier and a systematic low bias is seen in the lower panels for mooring 2.

Finally, we assessed the effect of reducing vertical resolution, in particular using one thermistor every $20\,\mathrm{m}$ {when computing $L_E$, but using instead information from all the thermistors for $L_T$} (case ``res'', crosses in figure \ref{fig:Subsampling}).
The agreement is very good for mooring 1, while a tendency to underestimate the results from the full data set is observed in mooring 2.
The limited impact of a reduction in vertical resolution is connected with the fact that the stratification is approximately constant on the scale of the mooring.

Overall, we can conclude that the method is robust to reductions in vertical and temporal resolution.
However, this positive result must not be overstated; the robustness of the ensemble average to changes in resolution holds only if the data set size is large enough to allow statistical convergence.
The method is particularly robust for mooring 1, which spans a longer period of time and measures stronger turbulence and larger overturns than {mooring 2}.
In {the latter}, on the other hand, the overturns are sparser in the time--depth plane and smaller, leading to the underestimation of the average turbulence level when variance is computed on a more limited portion of the water column.

\section{Discussion}
\label{sec:Discussion}

The results presented in the previous sections demonstrate that the analysis of temperature variance is a viable alternative to Thorpe scales analysis when sufficiently high time resolution is available.
In the best case scenario, which is represented here by mooring 1, the results from the MODWT can provide a quantitative estimate of $L_T$ and of $\epsilon_{L_T}$ over a wide range of values, on average approximately to within a factor of 2.
Even for more challenging data sets, however, as those recorded at moorings 2 and 3 (lower $SNR$, larger spread and lower values of $N$), the analysis technique described here can provide valuable information.
Furthermore, the technique can reliably distinguish phases with stronger turbulence from phases with weaker one at a mooring location, even when the quantitative agreement is lost.
In other words, linear correlation between the two estimates is present even if the slope of a linear fit is markedly smaller than 1 (e.g.~mooring 2) or even if a constant shift is present (e.g.~for $J\ne 5$ in mooring 1).

It must be stressed here that the correlation between $\epsilon_{L_E}$ and $\epsilon_{L_T}$ is for the largest part due to the correlation between the temperature variance and $L_T$.
In other words, correlation between $\epsilon_{L_E}$ and $\epsilon_{L_T}$ is not a consequence of using the same $N$ in the computation of the two estimates.
In fact, $N$ is only weakly correlated (order of $0.1$) with $\epsilon_{L_T}$, while $\Sigma_{J,t}^{M}$ has a much higher correlation with $\epsilon_{L_T}$ (order of $0.6$).
Only if the average over the whole time series at each depth is considered separately, $N$ is correlated with the turbulence dissipation rate, as deeper portions of the water column, closer to the bottom boundary layer, experience on average weaker stratification and stronger turbulence \citep[as discussed for instance in][their figure 25]{alford_observations_2000}.

{To identify the sources of error in equation \eqref{eq:ellison}, we first consider the propagation of the uncertainty of the temperature measurement.
The relative uncertainty of equation \eqref{eq:ellison} can be written as:
\[
\frac{\varepsilon_{L_E}}{L_E}\approx\sqrt{\frac{\varepsilon_{\overline{\theta'^2}^{1/2}}^2}{\overline{\theta'^2}} + \left(\frac{\varepsilon_{\partial \overline{\theta}/\partial z}}{\left|\partial \overline{\theta}/\partial z\right|}\right)^2},
\]
 where $\varepsilon_x/|x|$ represents the relative uncertainty of $x$, and where the covariance terms are neglected.
The first term on the right hand side can be interpreted as the squared inverse of the $SNR$ of the thermistors.
Considering the second term, we note that the uncertainty of $\partial \overline{\theta}/\partial z$ can be written as $\varepsilon_{\partial \overline{\theta}/\partial z}\left/\left|\partial \overline{\theta}/\partial z\right|\right.=\varepsilon_{\theta}\left/(\sqrt{2}\Delta z \left|\partial \overline{\theta}/\partial z\right|)\right.$, with $\Delta z$ the vertical spacing of the thermistors, whose uncertainty is neglected, and taking $\varepsilon_{\overline{\theta}}=\varepsilon_\theta$, surely an overestimate.
Covariance terms are again neglected, and the uncertainty is assumed to be a constant for all thermistors.
The factor $1/\sqrt{2}$ is due to the use of centered differences in the computation of the gradient.
Using the definition of $N_\theta$, the uncertainty of $L_E$ can be written as:}
\begin{equation}
  \frac{\varepsilon_{L_E}}{L_E} \approx \sqrt{\frac{1}{SNR^2} + \frac{g^2 \alpha^2}{2 \Delta z^2} \left(\frac{\varepsilon_{\theta}}{\left|N_{\theta}^2\right|}\right)^2}.
  \label{eq:error}
\end{equation}
Considering the sensor noise as the largest source of error, a $SNR$ for temperature better than $10^2$ (for the high frequency variations) should be sufficient to have a precise estimation of $L_E$ if only the first term on the right hand side of (\ref{eq:error}) is taken into account.
Considering figure \ref{fig:MODWT}, we see that at mooring 1, typical variations of temperature at periods shorter than $15\,\mathrm{min}$ are of the order of $10^{-2}\,^\circ\mathrm{C}$ (as confirmed by considering the MODWT decomposition of the whole data set; note that figure \ref{fig:MODWT} shows the variance rather than the RMS of the time series).
The $SNR$ is thus well above $10^2$ in mooring 1, having a typical noise level of $5\cdot10^{-5}\,^\circ\mathrm{C}$.
{The other mooring represents} a less optimal case, mainly due to the smaller temperature variations, closer to $10^{-3}\,^\circ\mathrm{C}$ in the frequency range considered.
Using mean values from the data, the second term on the right hand side of (\ref{eq:error}) is the main contribution to the relative error, being {approximately} $10\%$.
The relative error grows as $N_\theta^{-2}$ for weak stratification, and estimates with weaker stratification are thus more affected, likely one of the reason for the worse correlation observed for {mooring 2}.
However, in view of this error analysis which does not explain the observed dispersion, we suggest that the main source of error in the results is in fact the non--uniform stratification.
In other words, (\ref{eq:ellison}) is a linear approximation, as it translates temperature fluctuations into a length scale by means of the first derivative of the temperature profile.
In presence of finite displacements, changes in stratification will lead to inconsistencies between the temperature fluctuation and the estimated length scale.

The method described here may have an important practical application.
If only temperature measurements are available, a non-tight temperature-salinity relationship may produce inversions in the temperature profile which do not correspond to a density inversion.
This problem is likely less limiting for the method presented here.
By estimating the overturns length scale through Eulerian measurements, i.e.~measurements at a fixed depth, this method may provide an alternative to Thorpe scales estimation if sufficient temporal resolution is available.
Using a value of $N$ from, e.g.,~climatological data, the turbulence dissipation rate, or at least its variations in time with respect to a reference level, may be estimated.
This is particularly interesting considering the high correlation between the temperature variance and $\epsilon_{L_T}$ mentioned above.

Finally, it is worth discussing the link between our results and the classical problem of fine structure contamination, as first discussed by \citet{phillips_spectra_1971}.
In this context, fine structure contamination describes the leakage of power towards high frequencies of an Eulerian temperature signal in a layered medium.
Fine structure contamination {hinders} the computation of vertical displacement spectra from Eulerian measurements of temperature in layered media.
This problem, well known in the literature, is thought to render the temperature frequency spectrum above the internal wave unusable \citep{garrett_internal_1971,siedler_fine-structure_1974,mckean_interpretation_1974,gostiaux_fine-structure_2012}.
{Fine structure contamination is definitely present in the datasets presented here}, since layers are present in the temperature profiles, {in particular at mooring 1}.
However, the results show that there still is precious information available in the high frequency part of the temperature spectrum, despite fine structure contamination.

\section{Summary and conclusions}
\label{sec:Summary}

We discussed different ways of estimating overturning length scales from records of temperature.
We outlined a method for estimating the scales of overturns using frequency spectra of temperature beyond the internal wave band, using Eulerian measurements from moored thermistors.
We presented {two} recent oceanographic data sets of temperature, recorded close to the bottom at different slopes of Seamount Josephine (North Eastern Atlantic Ocean).
We found strong correlation between Thorpe and Ellison scales in these observations, similarly to previous numerical and laboratory experiments.
We pointed out the importance of a time--frequency decomposition of variance in order to find a good correlation between Thorpe and Ellison scales in oceanographic data sets.
In particular, we observed the highest correlation when frequencies beyond the internal wave band are considered, and we linked this to the loss of autocorrelation of the vertical temperature profile, probably due to diabatic process.
Finally, we discussed {that} these results suggest a possible use for records that suffer from fine structure contamination.

The {method presented} is particularly interesting if sufficiently long time series are available, from which the correlation between mesoscale dynamics and turbulence can be inferred, a subject that we aim to explore in detail in the future.
Further work is also needed in order to understand if the method can be applied in regions far from the bottom, with weaker and more intermittent turbulence.
Further work will also be devoted to the analysis of the temperature fluctuations at the highest frequencies recorded in the data, which we removed here (by subsampling the time series), their link to turbulence dissipation rate and their intermittency in space and time.

As a final important remark, we note that the results leave open the issue of the accuracy of the estimates in comparison to those from shear measurements using microstructure profilers.
Such a comparison would be particularly interesting considering that $\epsilon_{L_T}$ is likely overestimating the real turbulence dissipation rate during the initial phase of the turbulence evolution and possibly underestimating it during the late stage of turbulence decay, as suggested by the numerical simulations of \citet{smyth_length_2000} and \citet{mater_relevance_2013}.
If this is true, it may be worth considering $\epsilon_{L_E}$ for estimating turbulence dissipation rate during these phases.
In this respect, the present work may contribute to a better understanding of the distribution of turbulence dissipation rate in the ocean.

\begin{acknowledgments}
  The data used in this study is available at the Royal Netherlands Institute for Sea Research.

  The authors would like to thank the crew of RV~Pelagia for enabling the deployment and recovery of the instruments, and all the technicians, in particular Martin Laan, indispensable for the success of the measurements.

  The authors would like to thank the reviewers for their careful reading of the original manuscript, and for their detailed, constructive comments.
\end{acknowledgments}

\appendix
\section{}\label{sec:appA}

Here, we shortly present the MODWT decomposition method following the summary given in \citet{cornish_maximal_2006}.
For further insight into the theoretical and practical aspects of the MODWT, we refer to \citet{percival_wavelet_2006}, in particular Chapters 5 and 8.
Consider the time series of one thermistor at a fixed depth, $\theta(t)$, whose value at time $t$ is denoted $\Theta_t$.
Given a finite length time series $\{\Theta_t\}$ with unit time step, sampled at times $t=0,1,\ldots,(K-1)$, we want to decompose it in $J_0$ time scales $\tau = 2^{j-1}$, with $j=1,\ldots,J_0$.
The value of $J_0$ is limited by the length of the time series to values smaller than $\log_2 K$ i.e.~time scales shorter than $K/2$.
This is not a constraint in our analyses, given the length of the time series and the relatively low $J_0$ used.
At each level $j$ {and time $t$}, the decomposition implies the application of a wavelet filter ($\{h_{j,l}\}$, high-pass) and a scaling filter ($\{g_{j,l}\}$, low-pass), providing a set of wavelet ($\{W_{j,t}\}$) and scaling ($\{V_{j,t}\}$) coefficients respectively.
In their simplest form, $\{h_{j,l}\}$ and $\{g_{j,l}\}$ perform, respectively, a running differentiation and running mean of the time series.
The filters $\{h_{j,l}\}$ and $\{g_{j,l}\}$ can be computed at each level $j$ by stretching the $j=1$ base filters.
The length of the filters at each level is $L_j=(2^j-1)(L-1)+1$, with $L$ being the length of the base filter; {in other words, the index $l$ of $\{h_{j,l}\}$ and $\{g_{j,l}\}$ runs from $0$ to $L_j-1$}.
Using ``reflection boundary conditions'' to reduce the impact of the time series being of finite length and non-periodic, a new time series is defined from the original one:
\[
  \{\widetilde{\Theta}_t\} = \left\{
    \begin{array}{l l}
      \Theta_t & \quad \text{for } t=0,\ldots,K\\
      \Theta_{2K-1-t} & \quad \text{for } t=K,\ldots,2K-1.
    \end{array} \right.
\]
The wavelet and scaling coefficients are then defined iteratively for each $j$ as:
\begin{equation}
  \begin{split}
    W_{j,t} &= \sum_{l=0}^{L_j-1}h_{j,l}\widetilde{\Theta}_{j,t-l\bmod 2K}\\
    V_{j,t} &= \sum_{l=0}^{L_j-1}g_{j,l}\widetilde{\Theta}_{j,t-l\bmod 2K},
    \label{eq:MODWT}
  \end{split}
\end{equation}
where $\bmod$ represents the modulo operation.
{Equation (\ref{eq:MODWT}) is used first (at level $j=1$) on the original time series with ``reflection boundary conditions'', i.e.~$\{\widetilde{\Theta}_{1,t}\}=\{\widetilde{\Theta}_t\}$.
For higher $j$ levels, the procedure is iterated using the scaling coefficients obtained at the previous $j$ level, i.e.~$\{\widetilde{\Theta}_{j,t}\} = \{V_{j-1,t}\}$ for $j=2,\ldots,J_0$.
The residual of the decomposition is $\{V_{J_0,t}\}$, the scaling coefficients set at the maximum decomposition level.}
We stress that in practice boundary conditions are to a large extent irrelevant here, since we will consider short time scales compared to the time series length.
The MODWT wavelet and scaling coefficients have then been circularly shifted in time in order to align them with the original time series as described in \citet{percival_wavelet_2006} (page~112).
Each wavelet coefficient set thus obtained nominally characterizes the period range $[ 2^j, 2^{j+1})$, an octave, or in other words the time scale $2^{j-1}$.
The correspondence between MODWT level and period range is imperfect since the MODWT decomposition, as any other filtering procedure, suffers from spectral leakage, i.e.~leakage of power between nearby frequency bands.
The wavelet coefficients set at each level has zero average; together with the $J_0$-level scaling coefficients set they conserve the time series ``energy'':
\begin{equation}
  \|\{\Theta_t\}\|^2 = \sum_{j=1}^{J_0} \|\{\widetilde{W}_{j,t}\}\|^2 + \|\{\widetilde{V}_{J_0,t}\}\|^2,
  \label{eq:en-conserv}
\end{equation}
{where $\|{x_t}\|$ indicates $\sum_tx_t^2$ for a generic series $x_t$}.
From (\ref{eq:en-conserv}) a scale decomposition of the variance is obtained:
\begin{equation}
  \sigma^2_{\Theta_t} = \frac{1}{K}\|\{\Theta_t\}\|^2 - \overline{\Theta_t}^2 = \frac{1}{K}\sum_{j=1}^{J_0} \|\{\widetilde{W}_{j,t}\}\|^2 + \frac{1}{K}\|\{\widetilde{V}_{J_0,t}\}\|^2 - \overline{\Theta_t}^2,
  \label{eq:ANOVA}
\end{equation}
where $\overline{\Theta_t}$ is the mean value of $\{\Theta_t\}$.
The time--frequency decomposition of the variance we were looking for, at time $t$ and level $j$, is thus $\widetilde{W}_{j,t}^2$, whose expected value is the variance at the time scale $2^{j-1}$.
Since $\widetilde{W}_{j,t}^2$ is effectively a random process, rather than considering a single realization of $\widetilde{W}_{j,t}^2$ we will perform a non-overlapping running average of $M$ realizations to increase the statistical significance of the variance estimate:
\[
  \left<W^2\right>_{j,t}^{M} = \frac{1}{M}\sum_{l=t-(M-1)/2}^{t+(M-1)/2} \widetilde{W}_{j,l}^2,
\]
defined for times $t$ multiple of $M$ and for odd $M$.
Finally, we consider the sum over $J\le J_0$ levels:
\begin{equation}
  \Sigma_{J,t}^{M} = \sum_{j=1}^{J}\left<W^2\right>_{j,t}^{M},
  \label{eq:variance_estimate}
\end{equation}
which will be the key quantity used for computing $L_E$.
$\Sigma_{J,t}^{M}$ represents the variance in the temperature time series at periods shorter than $\tau=2^{J+1}$, {and by changing $J$ the variance up to different time scales can be estimated}.

In order to apply the technique outlined so far on the time series from each thermistor, the full data are first low-pass filtered and subsampled to a time step of $7\,\mathrm{s}$ using a non overlapping moving average {to speed up the computation.
The maximum level of the decomposition ($J_0$) is 13, which corresponds to a maximum period of 32 hours.
The combination of a time step of $7\,\mathrm{s}$ with 13 decomposition levels guarantees that we include in the MODWT the whole internal wave frequency band, from fast gravity waves with typical period of $2 \pi N^{-1}$ to the inertial motions (the inertial period at the latitude of the moorings is approximately 20 hours).
Furthermore, this choice of time step guarantees that we separate the inertial and the semidiurnal lunar tidal motions (period 12.42 hours) in two different levels.}
The subsampled time series can at this point be used to compute the wavelet and scaling coefficients defined in (\ref{eq:MODWT}).
The computation is performed using the pyramid algorithm described in detail in \citet{percival_wavelet_2006}.
The software used for this processing is available at the address https://sourceforge.net/projects/wmtsa/, and is mainly a python translation of the Matlab code available from the authors of \citet{percival_wavelet_2006} at the address http://faculty.washington.edu/dbp/wmtsa.html.
The wavelet filter used is the ``least asymmetric'' one of length $L=8$.
The results presented here depend very weakly on the choice of the wavelet type; the length of the filter is chosen as a compromise between the temporal (short filter) and frequency (long filter) resolution, and also to minimize the influence of the boundaries for the longest time scales considered.
The need to circularly shift the coefficients is a further reason to use the least asymmetric wavelets, since the phase shift introduced by them is known exactly a priori.

\section{}\label{sec:appB}

\begin{figure}
   \noindent\includegraphics[width=0.49\textwidth]{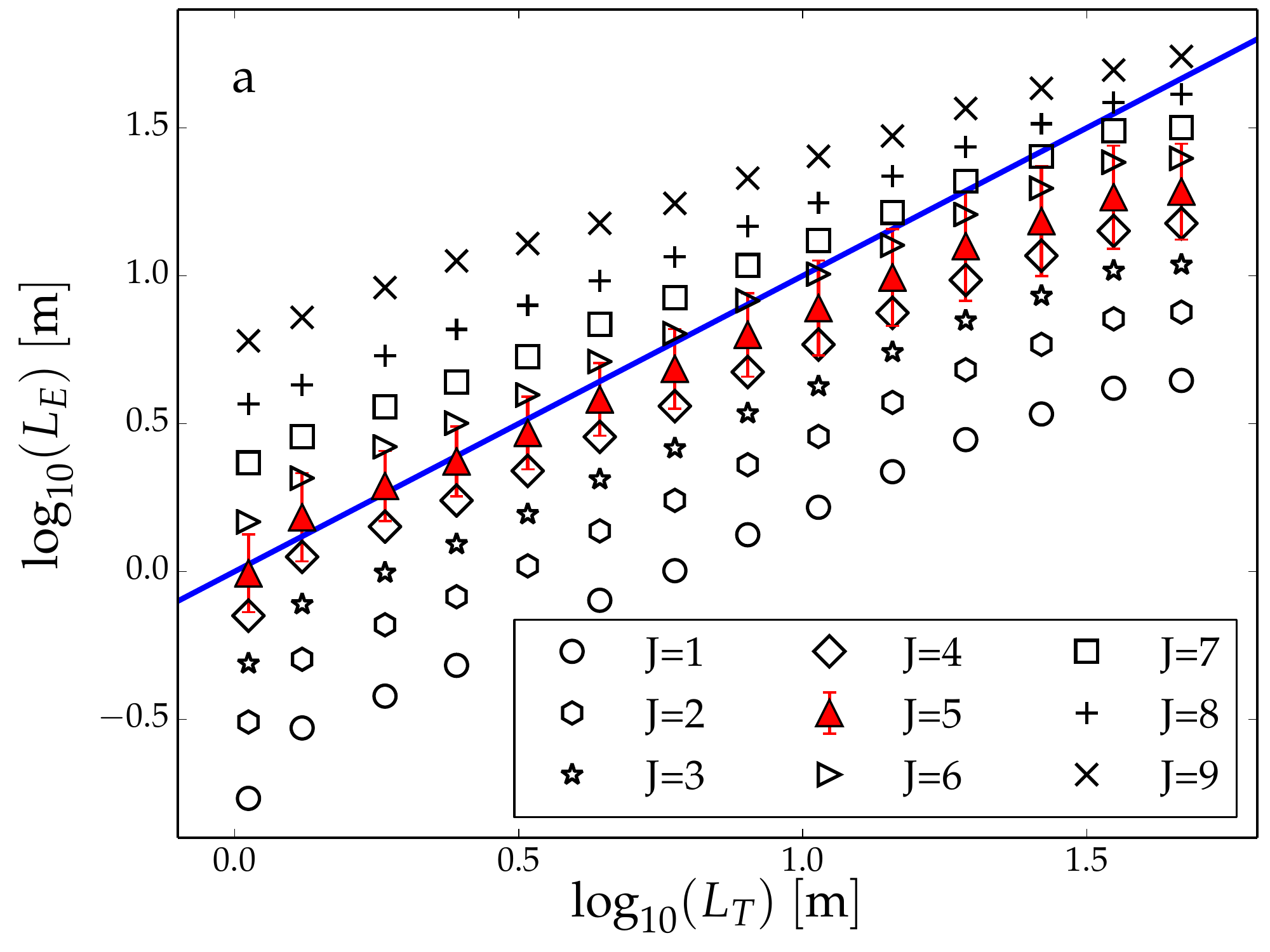}
   \noindent\includegraphics[width=0.49\textwidth]{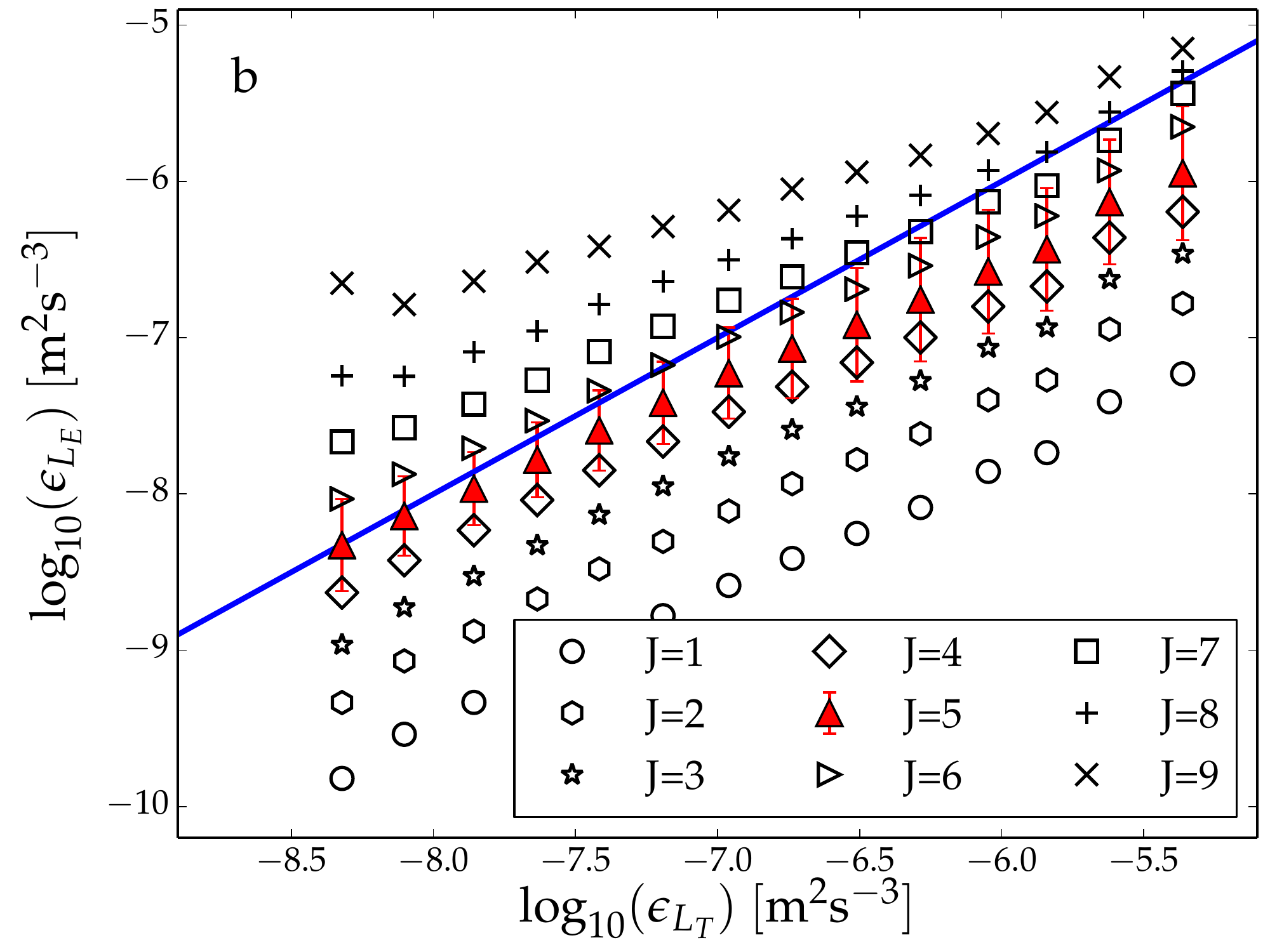}
  \caption{
    Similar to figure \ref{fig:LTLE1}, but showing only the median of the estimates with $J$ ranging from 1 to 9.
    The results for $J=5$, shown in greater detail in figure \ref{fig:LTLE1}, are in red and with an error bar equal to the RMS of the distribution.
  }
  \label{fig:Multiband}
\end{figure}

As discussed in section \ref{sec:Methods:Processing:Wavelet} and seen in figure \ref{fig:MODWT}-b, the spectrum of temperature variations is red.
As a consequence, the estimates (\ref{eq:LE_wavelet}) and (\ref{eq:eps_wavelet}) are most strongly influenced by the lowest frequencies included in the quantity $\Sigma_{J,t}^{M}$ (defined in equation \ref{eq:variance_estimate}).
In order to check how dependent the results are on the value of $J$ used, we show in figure \ref{fig:Multiband} an overview of the results using $J$ ranging from 1 to 7, for the data of mooring 1.
Figure \ref{fig:Multiband} shows that $J=5$ is a conservative choice which gives high correlation without overestimating the Thorpe scale based quantities.
However, values of $J$ ranging from 4 to at least 6 would still give a good agreement with the Thorpe scale based counterparts not only for what concerns correlation (see table \ref{tab:correlation}), but also for what concerns the order of magnitude.
This is even more valid if the RMS of the estimates is considered, shown in this figure only for $J=5$, and similar for the other values of $J$.
While the value of $N_{99}$ can provide a guide for choosing $J$, a conservative choice has to be preferred since the variations in time of $N$ and the unavoidable leakage of the filters will lead to the inclusion of some wave motions in the estimates for larger $J$.

\section{}\label{sec:appC}
{
As discussed in Appendix \ref{sec:appB}, the estimates (\ref{eq:LE_wavelet}) and (\ref{eq:eps_wavelet}) are most strongly influenced by the lowest frequencies included in the analysis.
The highest $J$ still giving good results is 6.
The largest period nominally included for this value of $J$ is $15\,\mathrm{min}$, and to resolve this time scale the minimal time step that can be used is $7.5\,\mathrm{min}$.
In principle, this should approximately be the minimal time step needed for obtaining a reasonable estimate of the Thorpe scale based quantities with the wavelet method.
However, this sampling rate also reduces the value of $M$ that can be used (number of successive variance estimates from MODWT averaged together), unless a strong reduction of the dynamic range of $L_E$ and $\epsilon_{L_E}$ is accepted.
Given these issues, it is found in practice that a sampling rate of $1\,\mathrm{min}$ is the minimum one providing results comparable to those shown in this work.
}


\clearpage

\begin{table}
  \label{tab:moorings}
  \centering
  \begin{tabular}{l l l l }
    \hline
    Mooring number & 1 & 2\\
    \hline
    Latitude & $36^\circ\, 58.885'\,\mathrm{N}$ & $37^\circ\, 01.431'\,\mathrm{N}$\\
    Longitude & $13^\circ\, 45.523'\,\mathrm{W}$ & $13^\circ\, 39.479'\,\mathrm{W}$\\
    Deepest thermistor & $2210\,\mathrm{m}$ & $2937\,\mathrm{m}$ \\
    Height above bottom & $5\,\mathrm{m}$ & $5\,\mathrm{m}$\\
    Number of thermistors & 144 & 140\\
    Thermistor spacing & $0.7\,\mathrm{m}$ & $1.0\,\mathrm{m}$\\
    Length & $100.1\,\mathrm{m}$ & $139\,\mathrm{m}$\\
    Deployment & 13 Apr 2013 & 12 Aug 2013\\
    Recovery & 12 Aug 2013 & 18 Oct 2013\\
    \hline
  \end{tabular}
  \caption[Details of moorings]{
    Details of the two data sets from the two moorings used.
    The length reported is the length over which thermistors are actually attached to the cable.
  }
\end{table}

\begin{table}
  \label{tab:correlation}
  \centering
  \begin{tabular}{l c c c c c}
    \hline
    $J$ & Max. period & $R[L_T,L_E]$ & $R[\epsilon_{L_T},\epsilon_{L_E}]$ & $R[L_T,L_E]$ & $R[\epsilon_{L_T},\epsilon_{L_E}]$\\
    & & \multicolumn{2}{c}{Mooring 1} & \multicolumn{2}{c}{Mooring 2} \\
    \cline{3-6} \\
    1  & $28\,\mathrm{s}$   & $0.82$ & $0.73$ & $0.73$ & $0.55$ \\
    2  & $56\,\mathrm{s}$   & $0.83$ & $0.74$ & $0.73$ & $0.55$ \\
    3  & $1.9\,\mathrm{min}$ & $0.84$ & $0.76$& $0.72$ & $0.55$ \\
    4  & $3.7\,\mathrm{min}$ & $0.84$ & $0.76$& $0.72$ & $0.54$ \\
    5  & $7.5\,\mathrm{min}$ & $0.85$ & $0.77$& $0.71$ & $0.54$ \\
    6  & $15\,\mathrm{min}$  & $0.85$ & $0.77$& $0.71$ & $0.53$ \\
    7  & $30\,\mathrm{min}$  & $0.84$ & $0.75$& $0.70$ & $0.51$ \\
    8  & $60\,\mathrm{min}$  & $0.82$ & $0.71$& $0.66$ & $0.45$ \\
    9  & $2\,\mathrm{hour}$  & $0.75$ & $0.59$& $0.61$ & $0.38$ \\
    10 & $4\,\mathrm{hour}$  & $0.66$ & $0.47$& $0.61$ & $0.39$ \\
    11 & $8\,\mathrm{hour}$  & $0.62$ & $0.41$& $0.64$ & $0.43$ \\
    12 & $16\,\mathrm{hour}$ & $0.56$ & $0.35$& $0.60$ & $0.34$ \\
    13 & $32\,\mathrm{hour}$ & $0.57$ & $0.34$& $0.62$ & $0.36$ \\
    \hline
  \end{tabular}
  \caption[Cross-correlation]{Cross-correlation between $L_T$ and $L_E$, and between $\epsilon_{L_T}$ and $\epsilon_{L_E}$, in the two moorings.
    The $J$ reported in the first column is the maximum level used for computing $\Sigma^{M}_{J,t}$ in \eqref{eq:LE_wavelet} and \eqref{eq:eps_wavelet}.
    The second column reports the maximum period nominally included in $\Sigma^{M}_{J,t}$.
    In all cases, $M=101$.
    Taking into account the autocorrelation of the time series with the method of \citet{ebisuzaki_method_1997}, all the values of cross-correlation reported are above the 99\% significance level, thanks to the length of the time series.
  }
\end{table}

\begin{table}
  \label{tab:subsamp}
  \centering
  \begin{tabular}{ c c }
    \hline
    Name & Description \\
    \hline
    ref & Reference case, using full data. \\
    cen & Data only from the central $1/5$ of the mooring. \\
    sub & Data subsampled in time. \\
    cen sub & Central $1/5$ of the mooring, subsampled in time. \\
    res & Using only 1 thermistor every $20\,\mathrm{m}$ \\
    \hline
    
  \end{tabular}
  \caption[Subsampling tests]{
    Summary and naming convention for different kinds of subsampling used to test the robustness of the results.
    For more details, see text.}
\end{table}

\end{document}